\documentclass[useAMS,usenatbib,usegraphicx]{mn2e}
\usepackage{amssymb}
\usepackage{amsmath}
\usepackage{multirow}
\usepackage{booktabs}
\usepackage{subfig}
\usepackage{color}
\usepackage{fixltx2e}
\usepackage{lscape}
\usepackage{enumitem}
\usepackage{comment}

\newcommand{\eg}{e.\,g. }
\newcommand{\ie}{i.\,e., }
\newcommand{\etc}{etc.\ }
\newcommand{\etal}{{et al.~}}

\newcommand{\galcount}{\textsc{galnum} }
\newcommand{\Dev}{\texttt{deV}}
\newcommand{\Exp}{\texttt{Exp}}
\newcommand{\Ser}{\texttt{Ser}}
\newcommand{\DevExp}{\texttt{deV-Exp}}
\newcommand{\SerExp}{\texttt{Ser-Exp}}
\newcommand{\pymorph}{\textsc{PyMorph}}
\newcommand{\galfit}{\textsc{Galfit}}
\newcommand{\sextractor}{\textsc{SExtractor}}

\newcommand{\readatlasimage}{\textsc{readAtlasImages-v5\_4\_11}}
\newcommand{\fits}{\textsc{FITS}}
\newcommand{\cmodel}{\texttt{CModel}}
\newcommand{\model}{\texttt{Model}}

\newcommand{\meertcat}{M15}

\usepackage{appendix}
\usepackage{longtable}
\usepackage{url}
\usepackage{float}
\usepackage[T1]{fontenc}
\usepackage{aecompl} 
\usepackage{hyperref}

\title[Meert {\em g}-,{\em i}-bands]{A Catalogue of Two-Dimensional Photometric Decompositions 
in the SDSS-DR7 Spectroscopic Main Galaxy Sample: Extension to {\em g}- and {\em i}-Bands}
\author[Meert \etal{}]{Alan Meert,$^{1}$\thanks{E-mail:
ameert@physics.upenn.edu} 
Vinu Vikram,$^{1}$\thanks{E-mail: vvinuv@gmail.com} 
and Mariangela Bernardi$^{1}$\thanks{E-mail: bernardm@sas.upenn.edu} \\
$^{1}$Department of Physics and Astronomy, University of Pennsylvania, 
Philadelphia, PA 19104, USA\\}

\begin{document}

\date{Accepted 2015 October 22.  Received 2015 Octoer 6; in original form 2015 July 27}

\maketitle

\label{firstpage}

\begin{abstract}
 
We extend the catalogue of two-dimensional, 
PSF-corrected de Vacouleurs, S\'{e}rsic,
de Vacouleurs+Exponential, and S\'{e}rsic+Exponential fits of $\sim7\times10^5$
galaxies presented in \cite{meert2014} to include the {\em g}- and {\em i}-bands. 
Fits are analysed using the physically motivated flagging system presented in
the original text, making adjustments for the differing signal-to-noise when necessary. 
We compare the fits in each of the {\em g}-, {\em r}-, and {\em i}-bands. 
Fixed aperture magnitudes and colours are also provided for all galaxies.
The catalogues are available in electronic format.
\end{abstract}


\section{Introduction}

The study of the structural components of galaxies has contributed substantially to the understanding
of the formation and evolution of galaxies. The structural components of galaxies in the local Universe 
trace morphological galaxy type and many other galaxy parameters related
to both assembly and evolution of galaxies: colour, metallicity, gas fraction, \etc
Properties may also trace halo
size and galaxy environment and place constraints on $\Lambda$CDM cosmology
\citep[\eg][]{blanton2005,Bernardi2009,shankar10a,shankar10b, kravtsov2014}. 
However, careful estimation of structural parameters 
for large numbers of galaxies is required to test different 
formation and evolution models.

This paper extends the original {\em r}-band fits presented in \cite{meert2014} (hereafter \meertcat{}) to the {\em g}- and {\em i}-band
so that colour comparisons can be made and colour-dependent galaxy properties can be measured.
\meertcat{} presented a 
catalogue of two-dimensional, PSF-corrected de~Vacouleurs, S\'{e}rsic,
de~Vacouleurs+Exponential, and S\'{e}rsic+Exponential fits of $\sim7\times10^5$
spectroscopically selected galaxies drawn from the Sloan Digital Sky Survey DR7 \citep[SDSS DR7;][]{sdss_tech,DR7} . Fits are presented for
the SDSS {\em g}- and {\em i}-bands utilizing the fitting routine \galfit{} \citep[][]{galfit} and analysis
pipeline \pymorph{} \citep[][]{pymorph}.

There has been much recent work on improving photometric decomposition of galaxies 
\citep[\eg][]{gadotti09, simard11, Kelvin2011, Lackner2012, Haussler2012}. 
These papers take different approaches to fitting multi-band imaging. 
\cite{simard11} (hereafter S11) fits the {\em g}- and {\em r}-bands jointly, producing a single set of fitted galaxy parameters that are band-independent 
(with the exception of magnitude, which is allowed to differ across the bands). \cite{Lackner2012}  (hereafter LG12) fit to the {\em r}-band data only, 
then refit to the {\em g}- and {\em i}-bands, holding all fitting parameters fixed, except for magnitude. \cite{Haussler2012} instead fits all available 
bands simultaneously allowing fitted parameters to vary smoothly across the bands. Here, we fit all 
bands independently of each other, using no information from neighbouring bands to inform our fits. 

While using multi-wavelength data to constrain fitting provides a more stable fit, especially in bands with lower signl-to-noise, it is not without downsides. Requiring the same 
fitting parameters across multiple bands ignores the intrinsic wavelength dependence of galaxies, their subcomponents, and the populations of stars contained within the components. This 
bias can be partially mitigated by allowing the relative magnitudes of the components (or the B/T) to have a wavelength dependent behavior. However, other parameters such as bulge radius
or S\'{e}rsic can have wavelength dependence as well. Fitting using the methods of \cite{Haussler2012} futher mitigates these biases by allowing parameters to vary smoothly across the 
measured bands. This increases the stability of the fits in all the bands. However, this prescrption can introduce systematic biases when the true wavelength
dependence of the profile does not correspond to the assumed dependence.

In the fits presented here, we fit each band independently, which makes no assumption about the wavelength dependence of the fitting parameters. This comes at the expense of reducing the
total S/N of the fitted parameters. It also makes the reliability of the fits sensitive to the band being fit, since the S/N of the images (and the galaxy components) vary across the 
bands. This also makes fitting more susceptible to dust features, especially in the bluer bands. While we do not directly address dust features, we do examine the effects of bars on 
galaxy features, ans we caution the reader that dust features may bias disc fits.

The simulations presented in \cite{meert2013} (hereafter M13) are used as a benchmark for these fits.
M13 used simulated galaxies drawn from {\em r}-band images of galaxies in the \meertcat{} catalog to test the accuracy of the fitting process. It established 
uncertainties on fitting parameters and showed that the choice of cutout size and background estimation were appropriate 
for SDSS galaxies. We update these simulations with additional fits that account for the difference in signal-to-noise for the 
{\em g}- and {\em i}-bands.

The paper is organized as follows: Section~\ref{sec:data} briefly describes the \meertcat{} catalogue used in this work. 
Section~\ref{sec:fitting} summarizes the \pymorph{} fitting routine used in \meertcat{} and in this work. 
Section~\ref{sec:sn_conerns} describes the relative differences in the signal-to-noise (S/N) 
between the {\em r}-band and the {\em g}- and {\em i}-bands and the changes made to the flagging to account for this.
Section~\ref{sec:results} describes the results of extending the {\em r}-band \meertcat{} flagging to the {\em g}- and {\em i}-bands,
provides comparisons incorporating morphological information, and shows colour comparisons using aperture magnitudes.
Finally, Section~\ref{sec:conclusion} concludes the
paper with a summary of results and final remarks. The fits discussed in this paper and further 
recommendations for their use are available in electronic format as a public release.

\section{The Data} \label{sec:data}

\begin{figure*}
\begin{minipage}{\linewidth}
\begin{center}
\subfloat[ ]{
\includegraphics[width=0.3\columnwidth]{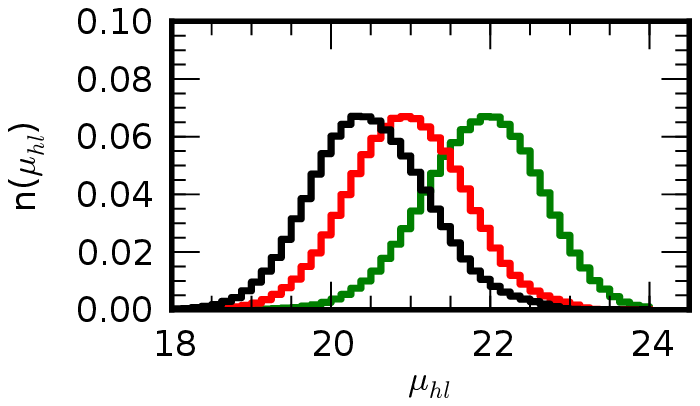}
\label{subfig:orig_dist:surfbright}}
\subfloat[ ]{
\includegraphics[width=0.3\columnwidth]{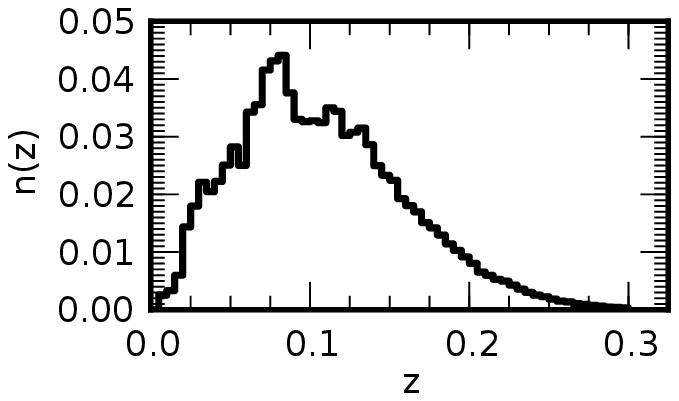}
\label{subfig:orig_dist:z}}
\subfloat[ ]{
\includegraphics[width=0.3\columnwidth]{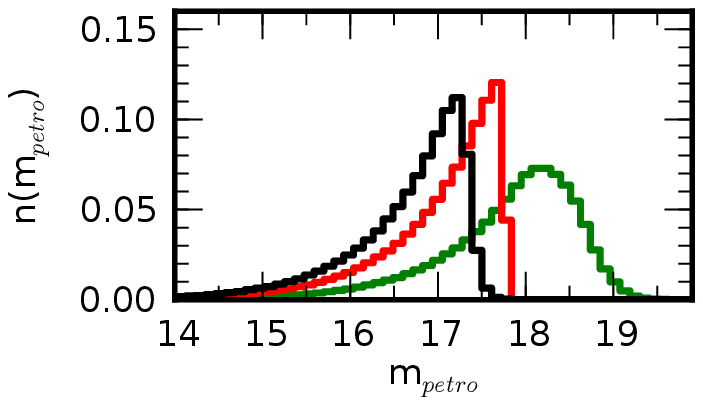}
\label{subfig:orig_dist:appmag}}
\end{center}
\end{minipage}

\begin{minipage}{\linewidth}
\begin{center}
\subfloat[ ]{
\includegraphics[width=0.3\columnwidth]{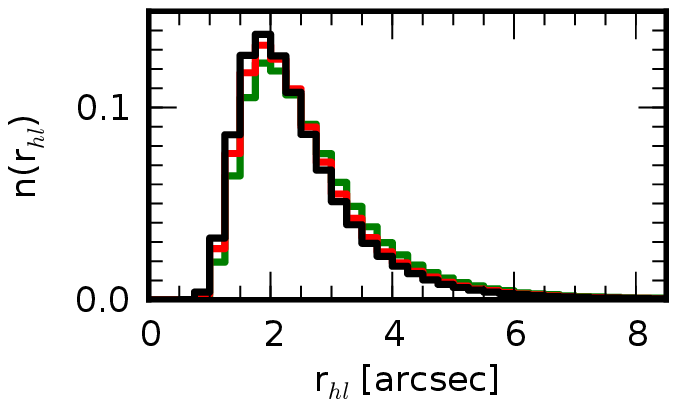}
\label{subfig:orig_dist:rad}}
\subfloat[ ]{
\includegraphics[width=0.3\columnwidth]{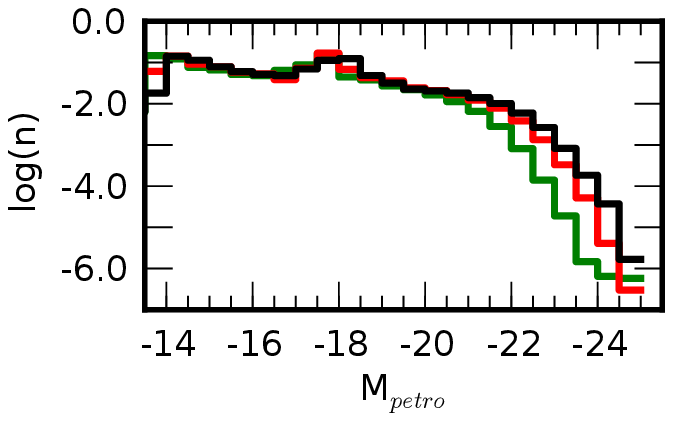}
\label{subfig:orig_dist:lum_func}}
\subfloat[ ]{
\includegraphics[width=0.3\columnwidth]{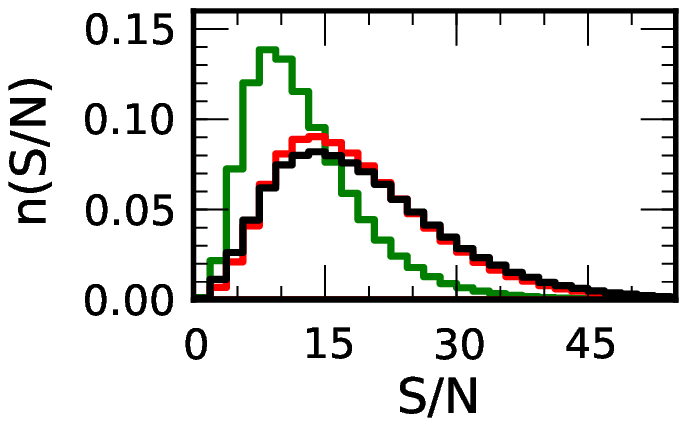}
\label{fig:SN}}
\end{center}
\end{minipage}

\caption{\textbf{\protect\subref{subfig:orig_dist:surfbright}}~The distibution of mean surface brightness within the halflight-radius,\textbf{\protect\subref{subfig:orig_dist:z}}~redshift distribution, \textbf{\protect\subref{subfig:orig_dist:appmag}}~extinction-corrected apparent Petrosian magnitude, 
\textbf{\protect\subref{subfig:orig_dist:rad}}~ Petrosian half-light radius, \textbf{\protect\subref{subfig:orig_dist:lum_func}}~ V$_{\mathrm{max}}$-weighted absolute Petrosian magnitude, and \textbf{\protect\subref{fig:SN}} Signal-to-Noise distribution using the the measurement of sky provided by the SDSS photometric pipeline of the sample used in this paper drawn from the DR7 SDSS spectroscopic galaxy sample. Bin counts are normalized {so that each distribution} integrates to 1. The {\em g}-, {\em r}-, and {\em i}-bands are shown separately as green, red, and black lines, respectively. The redshift distribution is 
band-independent and is only plotted in black.}
\label{fig:orig_dists}
\end{figure*}

The data used in this analysis are the same 670\,722 galaxies fit in \meertcat{}.
The \meertcat{} sample was drawn from the spectroscopic sample
of the Legacy area of the Sloan Digital
Sky Survey Data Release 7 \citep[hereafter SDSS DR7;][]{DR7}. 
The spectroscopic sample provides a
well-established sample with well-defined and tested selection
criteria. The criteria are presented in \cite{Strauss2002}. The primary 
selection criteria for these galaxies is that the extinction-corrected {\em r} band Petrosian magnitude between
magnitude 14 and 17.77. Thus the sample is magnitude limited and requires corrections in order to make robust conclusions about the 
population of galaxies.

Figure~\ref{fig:orig_dists} shows the distributions in the {\em g}-, {\em r}-, and {\em i}-bands using SDSS values
of mean surface brightness within the Petrosian halflight-radius, redshift, extinction-corrected aparrent Petrosian magnitude,
Petrosian half-light radius, V$_{\mathrm{max}}$-weighted
absolute Petrosian magnitude, and signal-to-noise for the sample used in
this paper.

The definition of the surface brightness and S/N used here is the same as in the previous paper, \meertcat{}. 
The mean surface brightness within the Petrosian halflight radius is defined as 
\begin{equation}\label{eq:avg_sb}
\begin{aligned}
\mu_{\textrm{50, r}}&=m_{petro,\ r} + 2.5\log{(2\pi r_{petro,\ 50}^2)}.\\
\end{aligned}
\end{equation}
We also define the S/N as the mean pixel flux within the half-light radius
divided by the noise associated with that pixel, or
\begin{equation}
\begin{aligned}
\left(\dfrac{\textrm{S}}{\textrm{N}}\right) &\equiv \dfrac{I_{\mu_{50}}}{N_{avg}}\\
N_{avg} &\equiv \sqrt{\dfrac{I_{\mu_{50}}+I_{sky}}{\textrm{gain}} + \textrm{dark variance}}\\
I_{\mu_{50}} &= 10^{-0.4*(\mu_{50}-zp)}*platescale^{-2}
\end{aligned}
\end{equation}
where $I_{\mu_{50}}$ is the source DN (`data numbers' or, equivalently, counts)\footnote{The counts
are related to the number of photo electrons collected by the detector through the gain of the detector amplifier. This distinction
is important since the photo electrons obey Poisson statistics.} of the average surface brightness defined in 
Equation~\ref{eq:avg_sb}. 
The zeropoint, $zp$, is calculated from the SDSS zeropoint, extinction, and airmass terms associated with each image.
The $platescale$ is used to convert the surface brightness from counts per square arcsecond to counts per pixel.
$N_{avg}$ is the noise in a pixel using the SDSS background
measurement as an estimate of the background flux and the average flux per pixel inside the Petrosian
half-light radius as the galaxy flux.

The surface brightness (top left panel) is brightest in the {\em i}-band and dimmest in the {\em g}-band. 
The sizes of galaxies in each band (bottom row, left panel) are similar, although the {\em g}-band is slightly larger than the 
{\em r}-band on average. The {\em i}-band is slightly smaller than the {\em r}-band. The large difference in surface brightness is due to substantial 
differences in the total magnitudes of the objects (top row, right panel). However, the S/N is most relevant to fitting. The S/N (bottom row, right) 
shows that the {\em i}-band has S/N similar to, and perhaps slightly higher than, the {\em r}-band. Although the {\em i}-band galaxy images are brighter and roughly the same 
size as the {\em r}-band galaxy images, which would imply a higher S/N, the sky brightness is also $\approx$50\% brighter in the {\em i}-band. 
The {\em g}-band has comparably lower average S/N relative to the {\em r}-band. The average {\em g}-band sky is dimmer by 1 magnitude relative to the {\em r}-band, 
but the galaxy brightness is also dimmer on average. Since the original simulations used to test the accuracy of the fitting algorithm
were based on {\em r}-band measurements, the true accuracy of fitting in the {\em g}-band may be substantially lower. This is most concerning in the case of the
\SerExp{} and \DevExp{} fits where two-components are being simultaneously fit, further lowering the S/N of each component (because the total flux is
now divided between the two fitted components). We will return to this issue in Sections~\ref{sec:sn_conerns} and \ref{sec:results}.

In addition to the data provided by CasJobs, \meertcat{} also used data from 
the Morphology catalogue of  \cite{huertas10} (hereafter referred to as H2011). 
H2011 is an automated morphological classification that used
a Bayesian SVM algorithm to morphologically classify all galaxies in the spectroscopic sample based on parameters 
in the SDSS DR7. K-corrections are calculated using version 4.2 of the K-correction code
\texttt{kcorrect} described in \cite{blantonKcorr}. To calculate the
K-correction, the SDSS \model{} magnitude and uncertainties are used and
data for all band passes (u,g,r,i,z) are provided to the program. 
We assume a cosmology with ($H_0$,$\Omega_\Lambda$,$\Omega_m$,$\Omega_k$) =(70 km s$^{-1}$Mpc$^{-1}$, 0.7, 0.3, 0.0) when necessary.
The original unique id numbers referred to as \galcount{} in \meertcat{} are also used throughout this text.

\section{The Fitting Process} \label{sec:fitting}

The fitting process is the same as in \meertcat{} with only minor modifications.  
FpC images and psField files are the primary data used in the fitting procedure. 
Postage stamp images of each source
were extracted from the fpC image such that the stamp was 40 Petrosian
half-light radii on a side (20*\texttt{petroR50\_r} from the centre
of the image to the edge) and centered on the target source.
In addition, a minimum size of 80 pixels on each side was set to ensure that enough
pixels were retained to properly determine the background.
We continue to set the cutout size based on the {\em r}-band sizes since the {\em r}-band data is used throughout SDSS to 
set consistent measures across all bands.

The band-specific PSF is extracted from the PsField files using the \readatlasimage{} program distributed on the SDSS 
site. \footnote{The use of \readatlasimage{} for PSF extraction is
described at \url{http://www.sdss.org/dr7/products/images/read_psf.html}}
The PSF provided by SDSS using the \readatlasimage{} program has a
standard image size of 51 pixels on each side.  The average PSF width for the 
bands is 1.47\arcsec, 1.35\arcsec, and 1.28\arcsec for the {\em g}-, {\em r}-, and {\em i}-bands, respectively.

Prior to fitting, the soft-bias is removed from the images
and PSF, we create sigma images from the SDSS images, and we normalize
the postage stamp and sigma images to a 1-second.

We again fit the \Dev{}, \Ser{},
\DevExp{}, and \SerExp{} models as in \meertcat{} to the sample described in
Section~\ref{sec:data} using \pymorph{} \citep{pymorph}. \pymorph{} is a Python based
automated software pipeline built on \sextractor{} \citep[][]{sex} and the two-dimensional fitting routine \galfit{} \citep{galfit}. 
The two-step neighbor identification process described in Section~3.4 of \meertcat{} 
is also applied to the {\em g}- and {\em i}-band fits.

\subsection{Measuring colours of fitted galaxies}\label{sec:tot_colour}
Galaxy colour is the ratio of light in two bands. Differences in the fitted parameters (\eg S\'{e}rsic index, radius) across the bands imply colour gradients. 
Analytical profiles integrated to infinitely large radius can have large portions of their total integrated light well beyond the apparent extent of the galaxies. 
For instance, a \Ser{} galaxy with n$=$4 has approximately 85 per cent of its total light within 4 half-light radii. For a \Ser{} galaxy with n$=$8, the total light contained within 4 half-light radii drops to approximately 76 per cent of the total galaxy light. Therefore, colour differences can be due to low levels of light at large radii. 
In order to mitigate this effect and make comparisons similar to common SDSS colours, we measure several aperture 
magnitudes that can be used for colour calculations.  

First, the fibre magnitude is measured for the 2 and 3\arcsec{} circular apertures representing the fibers used in BOSS \citep{dawson2013} and SDSS, respectively. These are measured using the PSF-convolved fitted model (\eg \Dev{}, \Ser{}, \DevExp{}, or \SerExp{}). The magnitudes can be used for comparison to the fibre magnitudes and fibre colours available in the SDSS database. 
In addition, magnitudes using the Petrosian aperture and our fitted models are measured. The Petrosian aperture is a circular aperture that is twice the {\em r}-band Petrosian radius.
This can be compared directly to the Petrosian magnitude and related colours. 

We also provide an aperture magnitude intended for comparison to the SDSS \cmodel{} and \model{} magnitude for each model and band. SDSS \Dev{} and  \Exp{} galaxy profiles fit by the \text{Photo} pipeline are truncated at 8$r_{dev}$ and 4$r_{exp}$, respectively. Therefore, we define the aperture radius as either 8$r_{dev, SDSS}$ or 4$r_{exp, SDSS}$ in the {\em r}-band depending on whether the SDSS \model{} magnitude uses the \Dev{} or \Exp{} fits.  This radius is used to define an elliptical aperture, from which we measure the aperture magnitude of each fitted model in all of the bands. In Section~\ref{sec:results} we compare the aperture colours calculated using these aperture magnitudes to the SDSS \model{} and \cmodel{} magnitudes.

\section{S/N concerns} \label{sec:sn_conerns}
\subsection{General effects of lower S/N}
Figure~\ref{fig:orig_dists} in Section~\ref{sec:data} showed that the S/N of the {\em i}-band is similar to that of the {\em r}-band 
originally studied in \meertcat{}. Therefore, we expect the performance of the flagging in the {\em i}-band to be similar to that of the 
{\em r}-band. However, the {\em g}-band fits were found to be lower in S/N with the peak of the distribution approximately a factor of 2 smaller than the peak in the {\em r}-band.

The simulations of M13 were primarily based on assuming the S/N of the {\em r}-band data, so further investigation may be necessary prior to accepting the {\em g}-band fits. M13 also tested fitting the SDSS using different S/N and resolution, focusing on cases of increased S/N and resolution that might be expected in other surveys, and showing that increasing resolution by a factor of 2 more strongly influenced the accuracy of the fitting than increasing S/N by a factor of 4 (see M13, Sections 3.3 and 3.4). 

According to M13, when the appropriate profile (\ie one vs. two-component) is chosen, 
magnitude and half-light radius are expected to be accurate at the 5\% level (using 1-$\sigma$ errorbars). However, these tests presumed the average S/N of an {\em r}-band image. This is not a concern for the {\em i}-band, which has both better seeing
and brighter galaxies. In the {\em i}-band the sky is also brighter offsetting any gains in S/N and making the distribution S/N very similar to the {\em r}-band, if not skewed slightly higher. In the {\em g}-band, the sky is dimmer, but the galaxies are also dimmer and the PSF 
wider such that the average S/N is substantially lower. We expect this to affect our measurements in a few ways. First, the fitting parameters, particularly in the case of the \DevExp{} and \SerExp{} fits may be biased or have very large uncertainty. Second, (and partially as a result of the first issue) the choice of bulge/two-component/disk based on the flagging may be biased.  

\begin{figure*}
\includegraphics[width=0.3\textwidth]{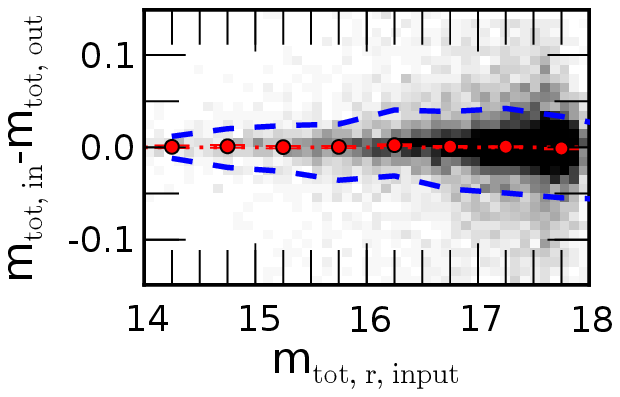}
\includegraphics[width=0.3\textwidth]{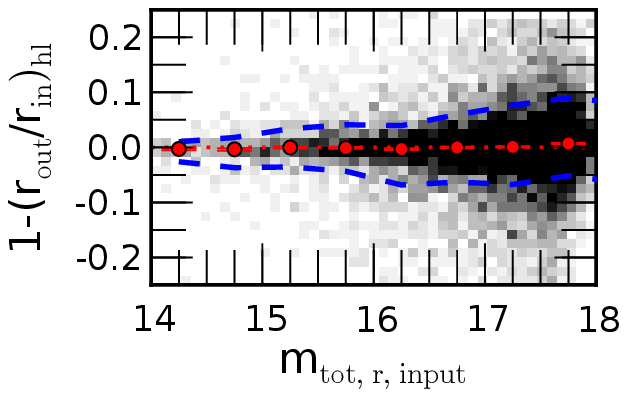}
\includegraphics[width=0.3\textwidth]{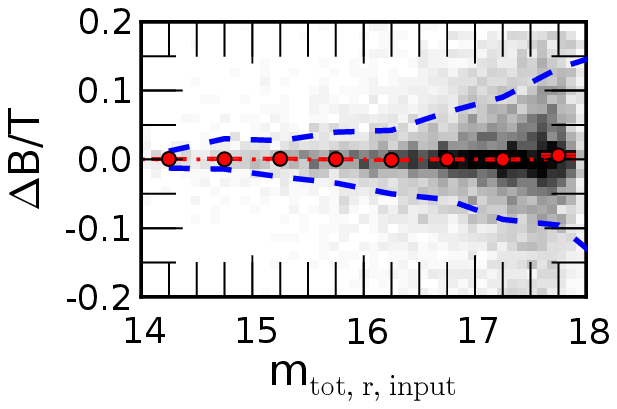}
\includegraphics[width=0.3\textwidth]{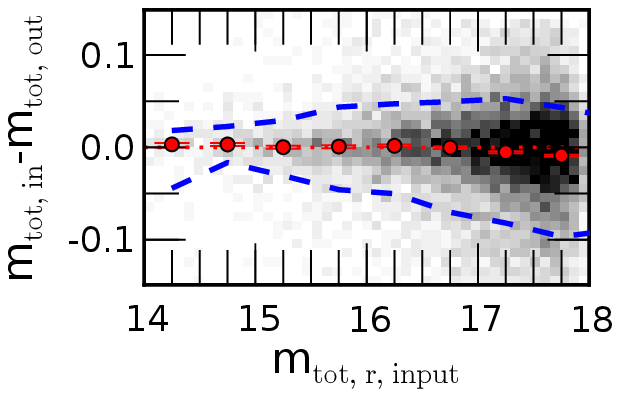}
\includegraphics[width=0.3\textwidth]{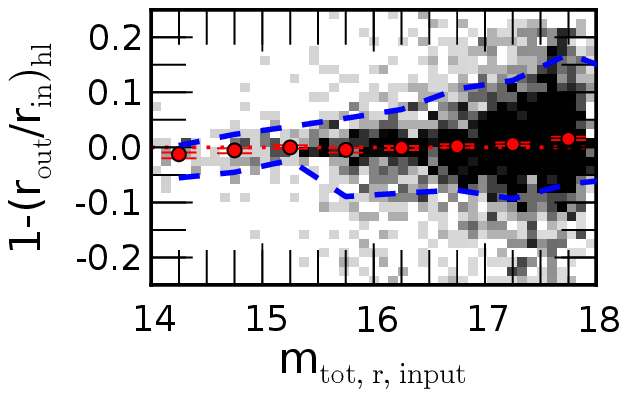}
\includegraphics[width=0.3\textwidth]{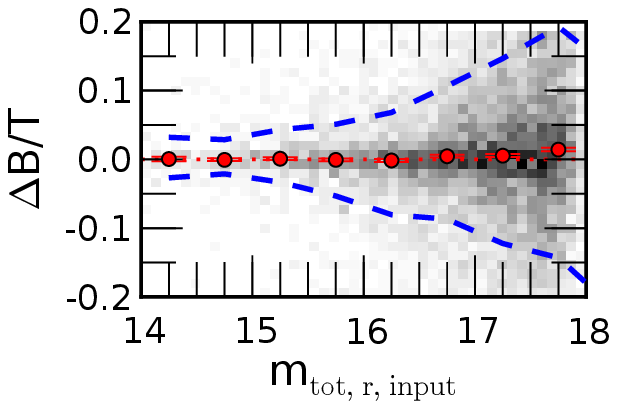}
\caption{Comparison of the original fiducial simulations of \SerExp{} fits published in M13 (top row) and the results of lowering the S/N by a factor of 2 (bottom row), similar to what we observe in the {\em g}-band images. Comparisons are shown for the total magnitude (left-hand column), total half-light radius (centre column), and B/T (right-hand column).
The median value is plotted in red with 95 per cent CI errorbars shown. The 68 percent scatter of the data is shown as the dashed contours. 
There is growth in the scatter of these estimated parameters by as much as a factor of two, 
but no additional bias is introduced beyond what was previously reported in M13. 
We expect the {\em g}-band estimates of magnitude and half-light radius to be accurate at the 15 per cent level (using 1-$\sigma$ errorbars) 
compared to 10 per cent accuracy in the {\em r}- and {\em i}-bands with biases similar to those reported in M13. 
The scatter grows most strongly at the dimmer magnitudes where S/N is already lowest.}\label{fig:serexp_sims}
\end{figure*}

Since our resolution is only moderately different between bands (less than 10 per cent) we expect the effects to be somewhat smaller than those reported in M13. However, prior to fitting the {\em g}-band, we simulated the M13 mock galaxies with S/N reduced by a factor of 2 to explore the effects of reduced S/N in the {\em g}-band. Figure~\ref{fig:serexp_sims} shows the behaviour of total magnitude (left-hand column), total half-light radius (centre column), and B/T (right-hand column) for the original fiducial simulations of \SerExp{} fits published in M13 (top row) and the results of lowering the S/N by a factor of 2 (bottom row), similar to what we observe in the {\em g}-band images. There is growth in the scatter of these estimated parameters by as much as a factor of two, but no additional bias is introduced beyond what was previously reported in M13. We expect the {\em g}-band estimates of magnitude and half-light radius to be accurate at the 15 per cent level (using 1-$\sigma$ errorbars) compared to 10 per cent accuracy in the {\em r}- and {\em i}-bands with biases similar to those reported in M13. The scatter grows most strongly at the dimmer magnitudes where S/N is already lowest. 

While they are not shown here, the sub-components of the \SerExp{} model are differently affected by the reduction in S/N. Disk components behave similar to the total magnitude and half-light radius with similar increases in scatter. Bulge components are more strongly affected by the change in S/N with a larger increase in the scatter, but no significant bias arising from the lower S/N. The difference between the bulge and disk components is likely due to the smaller number of free parameters in the disk versus the bulge (\ie the bulge S\'{e}rsic index adds
a substantial amount of freedom to the bulge model that the disk does not possess).

\subsection{Flagging adjustments as a result of S/N}

The flagging procedure presented in Section~4 of \meertcat{} is intended 
(among other goals) to reliably distinguish between one- and two-component fits, reducing any fitting bias resulting from fitting
a one-component profile to a two-component galaxy (see M13 for more information on the simulations and results). The majority of the 
flags are constructed using relative measurements that should not require adjustment to accommodate measurements in bands other than the {\em r}-band. For instance, determining that a galaxy has a contaminated bulge or disk component (flag bits 22 and 23) requires comparing the size of the bulge and disk components to the Petrosian radius in the corresponding band. 
Since the comparison is a relative one and does not have an absolute physical scale set, we expect the comparison to be valid across all the bands. 

However, the bounds for determining when a bulge or disk component is not present (\ie flag bits 1 and 4) are partially determined by 
a cut on magnitude calibrated to the {\em r}-band fits using a training set of visually classified galaxies. Since the magnitudes and S/N are different in the {\em g}- and {\em i}-bands, it is possible that these cuts need to be adjusted to work in the neighbouring bands.

In particular, the no bulge condition set in \meertcat{} was
\begin{equation}
 1000(B/T-0.8)^3 + (m_{disc}-19)>0.5
\end{equation}
and the no disc condition was 
\begin{equation}
 1000(0.2-B/T)^3 + (m_{bulge}-19)>0.5
\end{equation}

These cuts were set based on visual classification of fitted galaxies by the authors and the fact that scatter in the bulge or disc components grows substantially when the component was dimmer than 19. This cut of 19 was based on the assumption of S/N present in the {\em r}-band images. In particular, the average sky brightness and noise properties in the {\em g}-, {\em r}-, and {\em i}-bands differ as well as the average magnitudes. 

We computed the equivalent magnitude in the {\em g}-(and {\em i}-)bands of a component with the same S/N as an {\em r}-band magnitude 19.0 object while assuming median sky brightness, median bulge (or disc) half-light radius, and median noise properties. 
Table~\ref{tab:noise_properties} summarizes the median values used in each band and the equivalent magnitude for the {\em g}- and {\em i}-bands
calculated using these values. The {\em g}-band requires no adjustment. The {\em i}-band requires more adjustment due to the increasing sky brightness and dark variance in the band. We use these magnitudes as the basis for the flag bits 1 and 4 (\ie ``no bulge'' or ``no disk'') flags. These flags are the only flags that use an absolute cut on magnitude, so we expect that the remaining flags require no such adjustment. 

\begin{table*}
\begin{tabular}{c c c c c c c c}
 \textbf{Band} & \textbf{r$_{arcsec}$} & \textbf{zeropoint} &\textbf{$\mu_{\mathrm{sky}}$} [mag arcsec$^{-2}$] & \textbf{Gain} & \textbf{Dark Variance} & \textbf{S/N}& \textbf{m} \\
 g & 1.20 & 24.2 & 22.0 & 3.85 & 1.44 & 1.34 & 19.0\\
 r & 1.12 & 23.9 & 21.0 & 4.70 & 1.00 & 1.34 & 19.0 \\ 
 i & 1.00 & 23.6 & 20.3 & 4.86 & 5.75 & 1.34 & 18.2 \\ 
\end{tabular}
\caption{ Median values of radius, zeropoint, sky brightness, gain, and dark variance for the {\em g}-, {\em r}-, and {\em i}- bands. The S/N and magnitude are also shown. The {\em r}-band flags in \meertcat{} used a cut of 19 magnitudes as the basis for determining that a bulge or disk component is not reliable. Here, we show the median S/N of an object in the {\em r}-band, then using median values for the {\em g}- and {\em i}- bands, we calculate the equivalent magnitude of an object with the same S/N (within 1 per cent) as the {\em r}-band value. This band-dependent magnitude is used for bulge and disk flagging in the two-component {\em g}- and {\em i}-band fits and accounts for differences in the 
sky brightness and zeropoints between the bands.}\label{tab:noise_properties}
\end{table*}

\section{Results of Fitting and Flagging }\label{sec:results}

\begin{table*}
\begin{tabular}{l l l  p{2cm}  p{2cm}  p{2cm} }
\multicolumn{3}{l}{\textbf{Descriptive Category}} &  \textbf{\% serexp,\emph{g}} & \textbf{\% serexp,\emph{r}} & \textbf{\% serexp,\emph{i}} \\ \hline \hline
\multicolumn{3}{l}{\textbf{Trust Total and Component Magnitudes and Sizes}} & 33.290 & 39.055 & 37.049\\ \hline
& \multicolumn{2}{l}{\textbf{Two-Component Galaxies}} & 33.290 & 39.055 & 37.049 \\
 & & No Flags & 14.266 & 18.095 & 18.088 \\
 & & Good \Ser, Good \Exp\ (Some Flags) & 16.662 & 19.417 & 17.704 \\
 & &Flip Components, n$_{\Ser}<$2 & 2.362 & 1.543 & 1.258 \\ \hline
\multicolumn{3}{l}{\textbf{Trust Total Magnitudes and Sizes Only}} & 58.616 & 54.945 & 54.926\\ \hline
& \multicolumn{2}{l}{\textbf{Bulge Galaxies}} &  17.245 & 18.964 & 20.104\\
& &No \Exp\ Component, n$_{\Ser}>$2&  11.255 & 7.074 & 10.058 \\
& &\Ser\ Dominates Always &  5.990 & 11.890 & 10.046 \\
& \multicolumn{2}{l}{\textbf{Disk Galaxies}} &  31.197 & 25.146 & 24.193\\
& & No \Ser\ Component &   21.438 & 16.876 & 18.453\\
& & No \Exp, n$_{Ser}<$2, Flip Components &   1.455 & 0.551 & 0.744\\
& & \Ser\ Dominates Always, n$_{\Ser}<$2 &  0.121 & 0.103 & 0.074\\
& & \Exp\ Dominates Always &   2.013 & 2.872 & 1.607\\
& & Parallel Components &   6.170 & 4.745 & 3.316\\
& \multicolumn{2}{l}{\textbf{Problematic Two-Component Galaxies}} &  10.174 & 10.835 & 10.629\\
& & \Ser\ Outer Only &   7.218 & 7.504 & 8.383\\
& & \Exp\ Inner Only &   0.943 & 0.425 & 0.325\\
& & Good \Ser, Bad \Exp, B/T$>=$0.5 &   0.008 & 0.017 & 0.012 \\
& & Bad \Ser, Good \Exp, B/T$<$0.5 &   0.512 & 0.652 & 0.552 \\
& & Tiny Bulge, otherwise good &   1.493 & 2.237 & 1.357 \\ \hline \hline
\multicolumn{3}{l}{\textbf{Bad Total Magnitudes and Sizes}} &  8.094 & 6.000 & 8.025\\ \hline
& \multicolumn{2}{l}{Centering Problems} &  0.619 & 0.557 & 0.679 \\
& \multicolumn{2}{l}{\Ser\ Component Contamination by Neighbors or Sky} &  3.096 & 2.129 & 2.848 \\
& \multicolumn{2}{l}{\Exp\ Component Contamination by Neighbors or Sky} &  2.992 & 2.393 & 3.354 \\
& \multicolumn{2}{l}{Bad \Ser\ and Bad \Exp\ Components} &  0.824 & 0.239 & 0.347 \\
& \multicolumn{2}{l}{Galfit Failure} & 0.418 & 0.355 & 0.311 \\
& \multicolumn{2}{l}{Polluted or Fractured} & 1.038 & 0.676 & 0.946 \\
\end{tabular}
  \caption{A breakdown of the \meertcat{} flags now shown for the {\em g}-, {\em r}-, and {\em i}-band fits. The prevalence of galaxies with two-component  fits decreases toward the bluer end from $\approx$ 39 per cent in the {\em i}-band to 33 per cent in the {\em g}-band. The prevalence of pure bulge galaxies also decreases toward bluer bands while the prevalence of disk galaxies increases. This is as expected using a simple bulge+disk model with a bluer disk and redder bulge.}
\label{tab:serexp_flags}
\end{table*}

We again use the series of physically motivated flags analysed in detail in \meertcat{}.
We use these flags to determine the reliability of the various fits and the individual sub-components.
We also use the flags to mark poorly fitted galaxies. 
Table~\ref{tab:serexp_flags} shows the flagging categories, the final percentage of \SerExp{} galaxies in each flagging category
for the {\em g}-, {\em r}-, and {\em i}-bands (left, centre, and right columns, respectively).

\meertcat{} used a tiered structure to describe the fits. The most general description is the accuracy of the total magnitude and radius. \meertcat{} showed that about 94\% of the {\em r}-band \SerExp{} sample is classified as having an accurate measurement of the total magnitude and half-light radius. This percentage is reduced in the {\em g}- and {\em i}-bands to approximately 92 percent. The increase in failed fits can be mostly attributed to increasing neighbour contamination, which is reflected in the flagging as an increase of the ``\Ser{} component contamination by Neighbours or Sky'' and the  ``\Exp{} component contamination by Neighbours or Sky'' flags which
contain an additional $\approx$1.5 per cent of the total sample compared to the original {\em r}-band fits of \meertcat{}. This increase in neighbour contamination and failed fits is not due to the increase in number of neighbours (as these are the same set of galaxies in each band), but instead due to the decrease in the number of two-component galaxies. Fewer two
component galaxies means that more second fit components(either bulge or disc) will have little or no light to fit, and can be used by the fitting algorithm to reduce fit residuals by trying to fit light from neighbour galaxies. 

In addition to the increase of failed galaxies, there is a decrease in the percentage of two-component fits from $\approx$39 percent in the {\em r}-band to 33 percent in the {\em g}-band and 37 per cent in the {\em i}-band. At the same time, the number of pure bulge galaxies drops in the bluer filters from 20 percent in the {\em i}-band to 17 per cent in the {\em g}-band and the number of disks increases substantially from 24 per cent in the {\em i}-band to 31 per cent in the {\em g}-band. These observations agree with a simple bulge+disk model of galaxies where the disk tends to be bluer and the bulge is redder. It may also partially explain the increase in failed cases due to contamination or sky since the {\em g}-({\em i}-) band will feature the disk(bulge) more prominently and the second component will be dimmer or not present due to the different colour. With dim or low S/N component, there is a more likely possibility that the fitting routine will settle into a local minima that improperly uses this second component since it is not well constrained by the observations. A poorly constrained component may be used to fit stray light from neighbouring galaxies, increasing the failure rate as observed here. 

\subsection{Flagging versus basic fitting parameters}
\begin{figure*}
 \centering
\includegraphics[width = 0.95\linewidth]{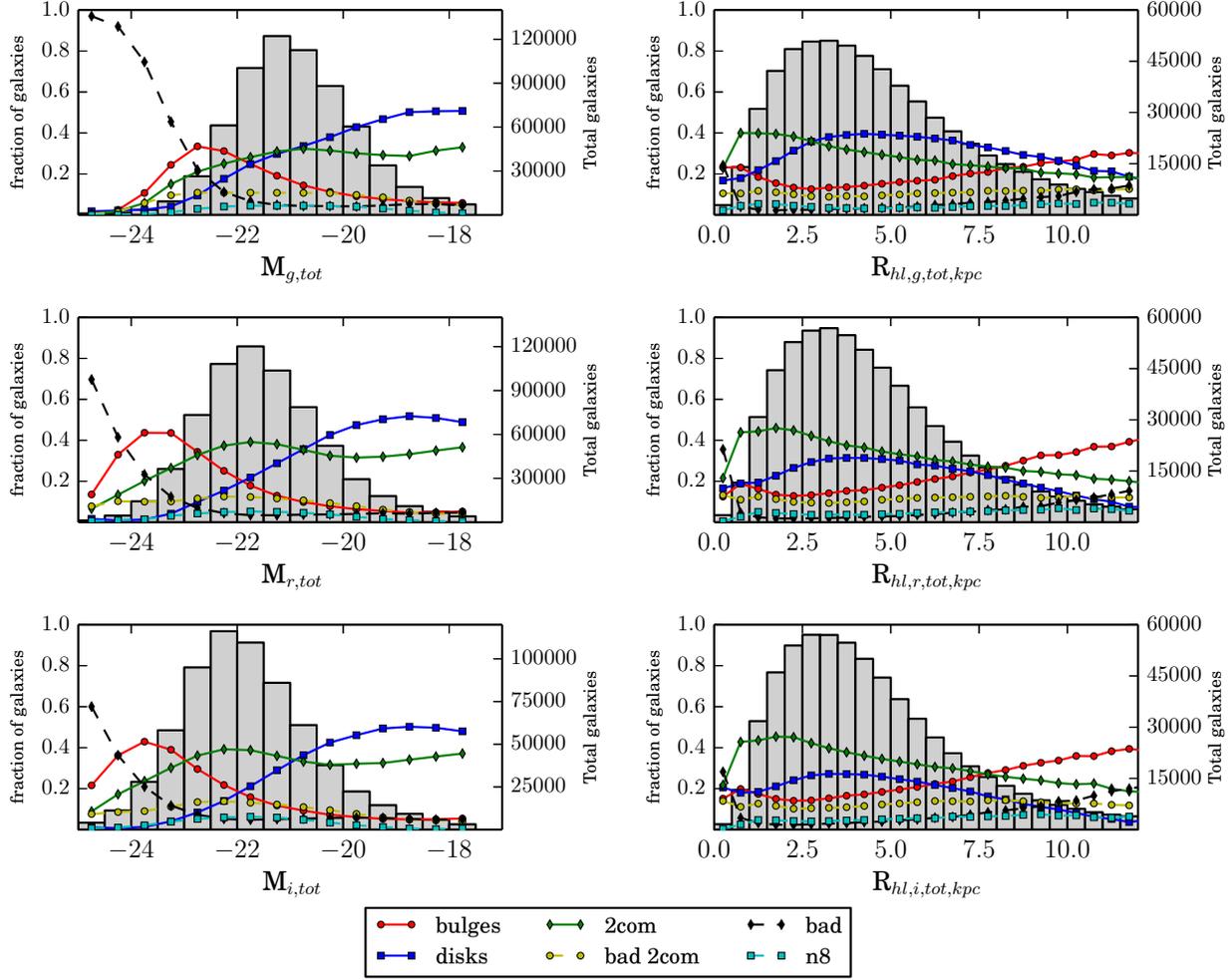}
\caption{The magnitude distribution of bin-by-bin percent of galaxies of each type; Disk (blue points, labelled as ``disks''), Bulge (red points, labelled as ``bulges''), Two-Component (green points, labelled as ``2com''), Problematic Two-Component (yellow points, labelled as ``bad 2com''), Failed Galaxies (black points, labelled as ``bad''), according to our categorical flags. We have also separated the good two-component models into those with S\'{e}rsic indices below 8 and those galaxies with acceptable fits but the  S\'{e}rsic index  of the bulge hits the $n=8$ boundary of the parameter space (cyan points, labelled as ``n8'') to check for any bias resulting from the restriction on the fitted S\'{e}rsic index. In the background of the plot, we plot the total distribution of galaxies with respect to the parameter used to bin the data. }
\label{fig:flag_by_type_gri_mag}
\end{figure*}

Figure~\ref{fig:flag_by_type_gri_mag} shows the absolute magnitude distribution (left-hand side) and the physical half-light radius distribution (right-hand side) of the {\em g}- (top row), {\em r}- (middle row), and {\em i}-band (bottom row) for galaxies in the catalogue. The figure shows the bin-by-bin percent of galaxies of each type; Disk (blue points, labelled as ``disks''), Bulge (red points, labelled as ``bulges''), Two-Component (green points, labeled as ``2com''), Problematic Two-Component (yellow points, labelled as ``bad 2com''), Failed Galaxies (black points, labelled as ``bad''), according to our categorical flags. We have also separated the good two-component models into those with S\'{e}rsic indices below 8 and those galaxies with acceptable fits but the  S\'{e}rsic index  of the bulge hits the $n=8$ boundary of the parameter space (cyan points, labeled as ``n8'') to check for any bias resulting from the restriction on the fitted S\'{e}rsic index. The background shows the total distribution of galaxies with respect to the parameter used to bin the data.

The fits of this work contain almost no bulges (only a few percent) and a mixture of disk and two-component galaxies at low magnitudes (below about -19.5). Two-component fits are the dominant model between -20 and -22 and bulges dominate at magnitudes brighter than -22. In the {\em g}-band, the failure rate is substantially higher at the brightest magnitudes (above -24 in the {\em g}-band). It is reasonable to expect that the majority of these bright {\em g}-band galaxies be incorrectly fit, dimmer galaxies farther than true bright {\em g}-band galaxies. The {\em g}-band sample is dimmer than the {\em r}-band and galaxies as bright as -24 in the {\em g}-band are extremely rare due to the surface-brightness, flux, and size limits of this sample, making it more likely that galaxies as bright as -24 are indeed incorrectly fit galaxies. 
|
The {\em r}- and {\em i}-bands appear similar in behaviour, while the {\em g}-band shows increased contributions from pure disk galaxies in the absolute magnitude distribution. 
In the {\em g}-band magnitude distribution (top left) the range of magnitudes with majority pure disk population is substantially larger compared to the {\em r}- and {\em i}-bands (middle and lower rows). The physical half-light radii are also similar in behaviour across the bands. This shows that the fitting
is quite similar in distribution across the bands with a smooth transition from having more disks in the {\em g}-band to fewer disks in the
{\em i}-band. However, in addition to stability across the entire sample, single galaxies should show strong correlation across the three
fits.

\begin{figure*}
\centering
\includegraphics[width=0.32\linewidth]{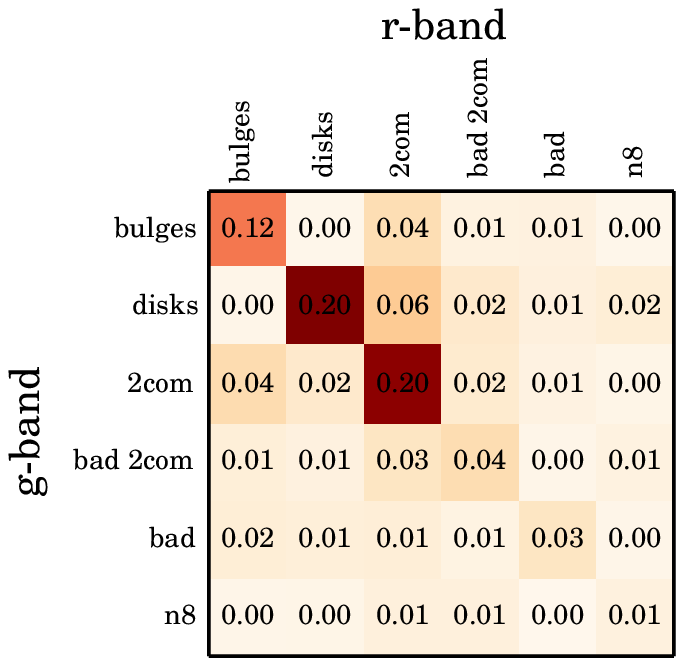}
\includegraphics[width=0.32\linewidth]{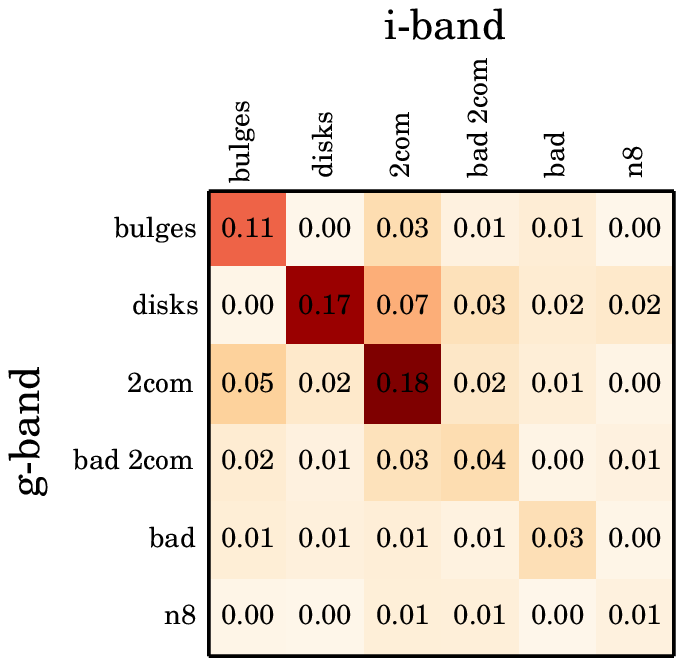}
\includegraphics[width=0.32\linewidth]{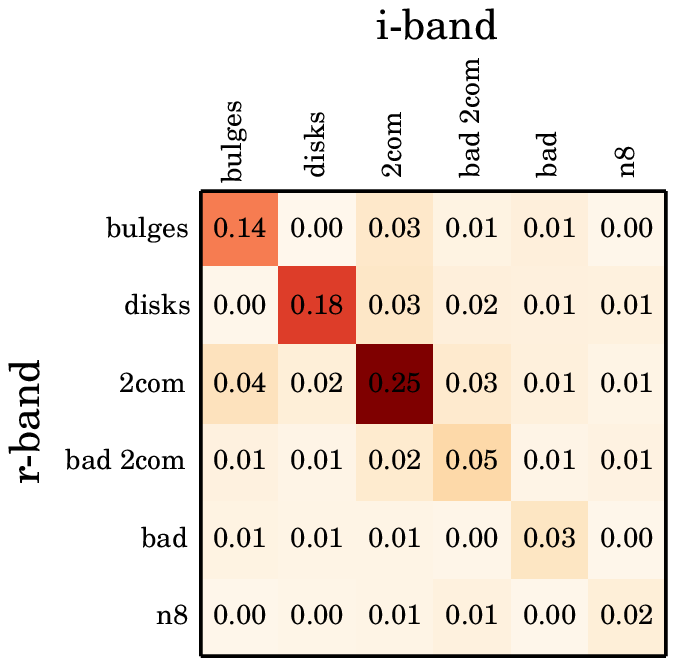}
\caption{The confusion matrix for each combination of bands and flagging categories. The number in each cell denotes the fraction of the catalogue classified by the combination of categories and bands denoted on the axes. For example, the top left cell of the first matrix shows that 12 percent of the sample is classified as ``pure bulge'' in both the {\em g}- and {\em r}-band fits. All possible combinations of bands are shown. The majority of galaxies ($\approx65-70$ per cent) are in the top left 3$\times$3 quadrant of each matrix, indicating that the majority of galaxies are similarly identified across the bands. Also, the bulge-disk and disk-bulge cells are below 0.5 per cent (indicated by the 0.00 value) showing that few or no galaxies switch directly from bulge to disk categories, which would be unexpected.}\label{fig:correlation_matrix}	
\end{figure*}

In Figure~\ref{fig:correlation_matrix} we show the breakdown of the fitting categories across two bands at a time. Each cell of the
matrix shows the fraction of the entire catalogue that is classified in the combination of bands and fitting categories denoted by the axes of the matrix (\eg the top left cell of the first matrix shows that 12 percent of the sample is classified as ``pure bulge'' in both the {\em g}- and {\em r}-band fits). The majority of galaxies ($\approx65-70$ per cent are in the top left 3$\times$3 quadrant of each matrix, indicating that the majority of galaxies are similarly identified across the bands. Also, the bulge-disk and disk-bulge cells are below 0.5 per cent (indicated by the 0.00 value) showing that few or no galaxies switch directly from bulge to disk categories, which would be unexpected. Additionally, the cells along the diagonal contain between 50 and 60 percent of the sample, indicating that the majority of galaxies receive the same broad classification in neighbouring bands. This demonstrates that there is 
stability in the flagging classification used in this work.

\subsection{Examining the colour}
As discussed in Section~\ref{sec:tot_colour}, we provide aperture magnitudes as part of this data release to allow colour measurements. In this section, we make some simple comparisons 
of colours to commonly used colours obtained from the SDSS database.

\meertcat{} assigned a T-type to each galaxy in the catalogue using the type probabilities (Ell, S0, Sab, Scd) provided by the BAC 
\citep[][]{huertas10} (see \meertcat{} for additional details). The BAC of \cite{huertas10} used a Bayesian approach to assign probabilities of being one of four broad galaxy types 
(Elliptical, S0, Sab, or Scd) to each galaxy using colour, total axis ratio, and concentration as measured by the 
SDSS. An SVM algorithm was used to produce the probabilities, using the \cite{fukugita2007} sample as the training set. 

\begin{figure*}
\centering
\includegraphics[width=0.32\linewidth]{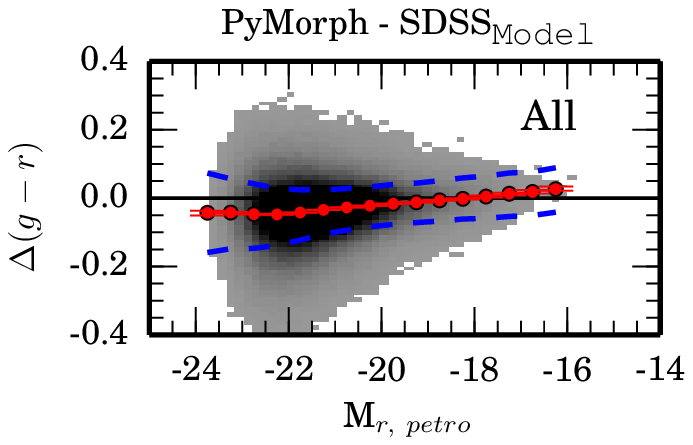}
\includegraphics[width=0.32\linewidth]{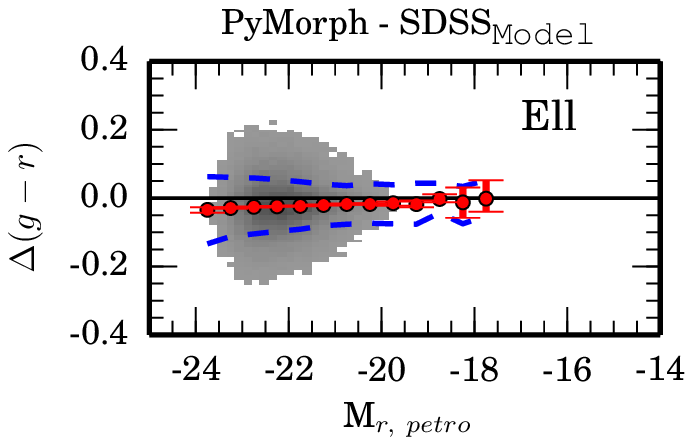}
\includegraphics[width=0.32\linewidth]{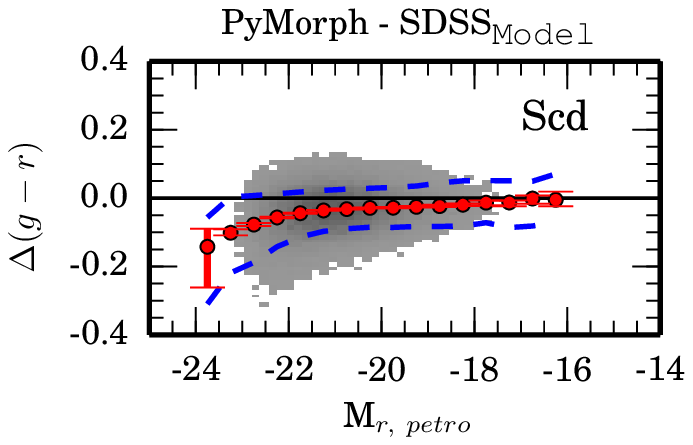}

\includegraphics[width=0.32\linewidth]{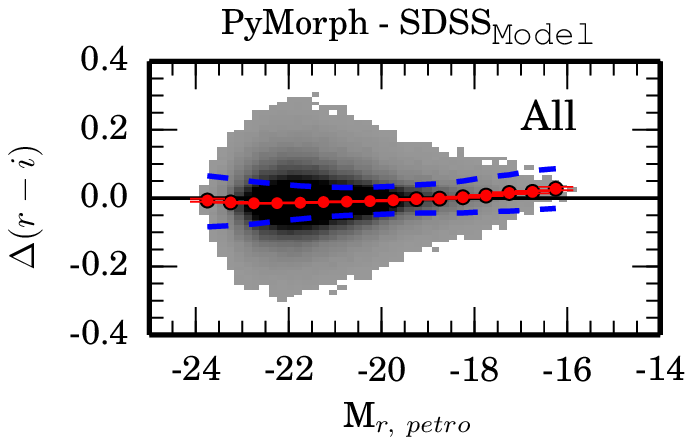}
\includegraphics[width=0.32\linewidth]{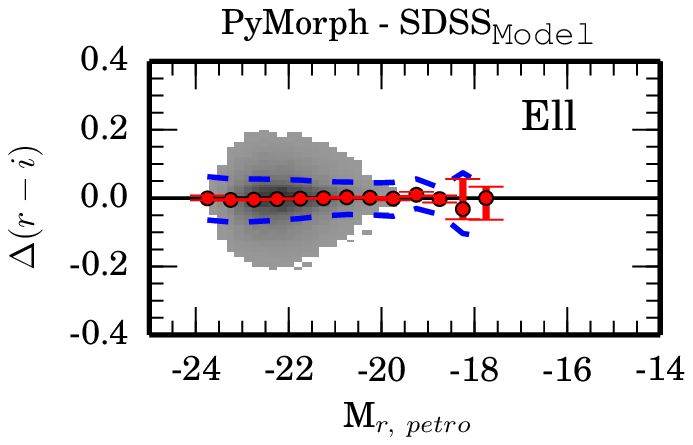}
\includegraphics[width=0.32\linewidth]{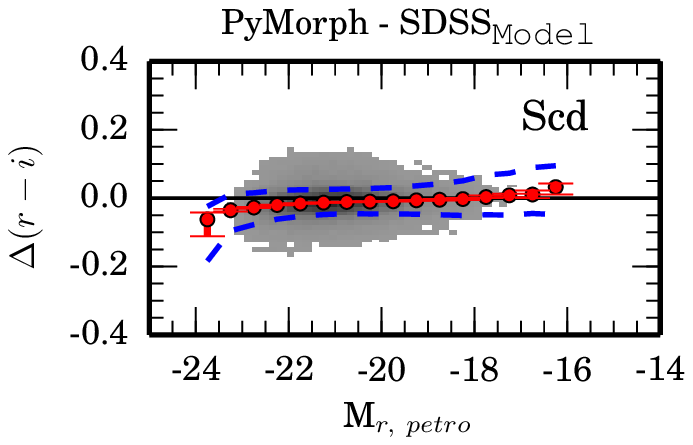}

\includegraphics[width=0.32\linewidth]{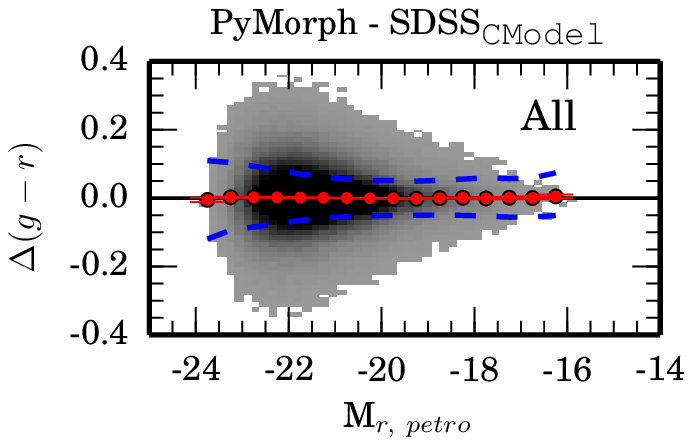}
\includegraphics[width=0.32\linewidth]{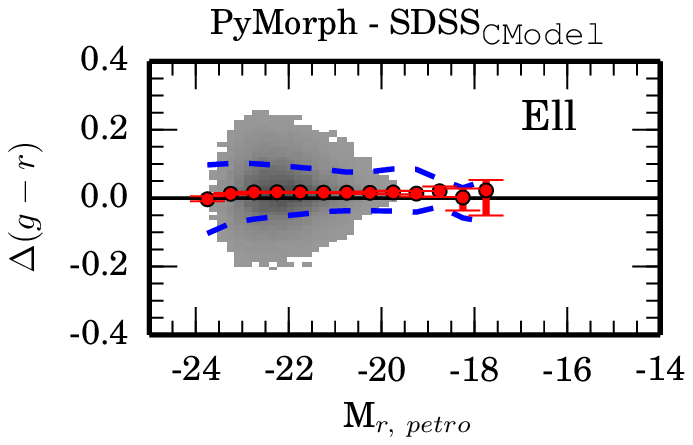}
\includegraphics[width=0.32\linewidth]{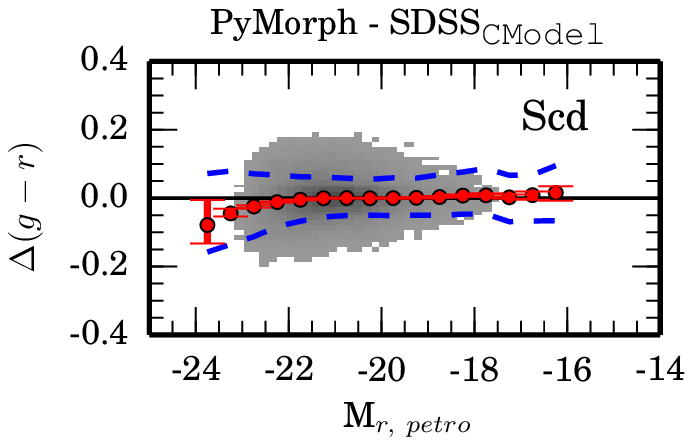}

\includegraphics[width=0.32\linewidth]{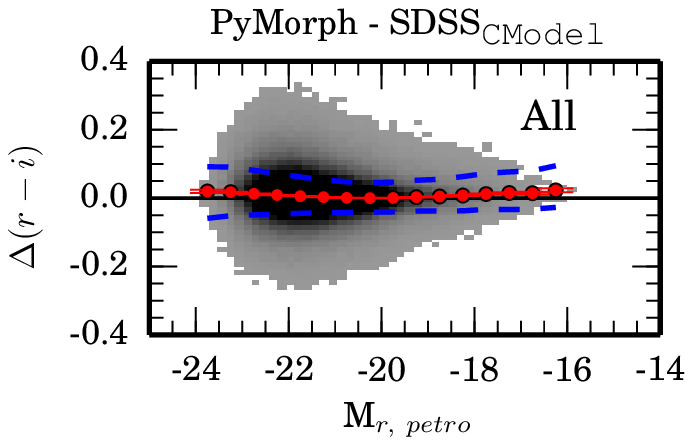}
\includegraphics[width=0.32\linewidth]{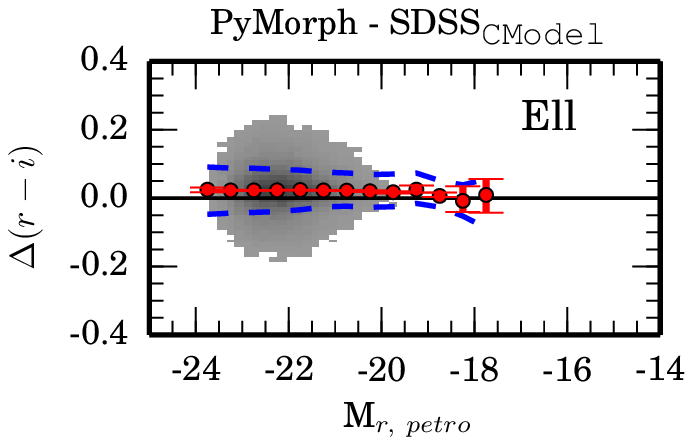}
\includegraphics[width=0.32\linewidth]{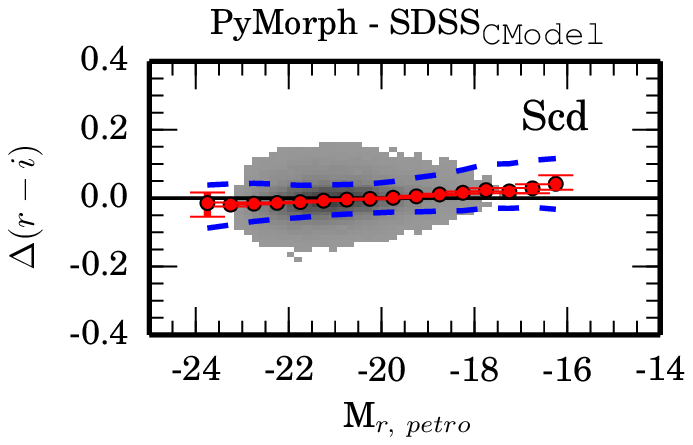}

\caption{\pymorph{} \SerExp{} minus SDSS \model{} colours (top two rows) and \pymorph{}  \SerExp{}  minus SDSS \cmodel{} colours (bottom two rows)
as a function of absolute Petrosian magnitude for the
{\em g}-{\em r} and {\em r}-{\em i} colors. The colours are shown for all galaxies, Ell, and Scd galaxies 
in the left-hand, centre, and right-hand columns, respectively. 
Median values are plotted in as red symbols with 95 per cent CI as error bars. The 68 per cent contours are shown as blue dashed lines. \pymorph{} colours appear 
similar to the SDSS \cmodel{} colours with some scatter, but little or no apparent trend with respect to magnitude.}\label{fig:color_cmp}	
\end{figure*}

Figure~\ref{fig:color_cmp} compares the \pymorph{} \SerExp{}, SDSS \model{}, and SDSS \cmodel{} colours as a function of absolute Petrosian magnitude for the full sample, Ell, and Scd galaxies in the left-hand, centre, and right-hand columns, respectively.
The {\em g}-{\em r} and {\em r}-{\em i} colours are shown in separate rows.  

When comparing the \pymorph{} \SerExp{} and SDSS \model{} colours, the {\em g}-{\em r} colour shows a slight negative bias ($\sim 0.025-0.05$) 
indicating that \pymorph{} \SerExp{} models are slightly bluer compared to 
SDSS \model{} colours for the total sample. When looking at only Elliptical galaxies, the bias is reduced. This is not the case for the Scd galaxies. 
The scatter is also broader in colours based on the {\em g}-band ($\approx 0.075$ vs $\approx 0.05$ in the {\em r}-{\em i} colour). 
When comparing \pymorph{} \SerExp{} and SDSS \cmodel{} colours, the bias disappears but the scatter remains relatively broad.
The SDSS \model{} fixes the fit type to be pure \Exp{} or pure \Dev{} for the galaxies. This static choice of model may play a role in the differences observed here. Fixing the model across the bands will certainly reduce the scatter in the colours, as the only way to adjust the model is an overall normalization (changing the total magnitude). However, as previously
stated, this ignores the changing profile of the galaxy across the bands as well as the associated colour gradients, which may contain interesting information about the galaxy. 

\begin{figure*}
\centering
\includegraphics[width=0.32\linewidth]{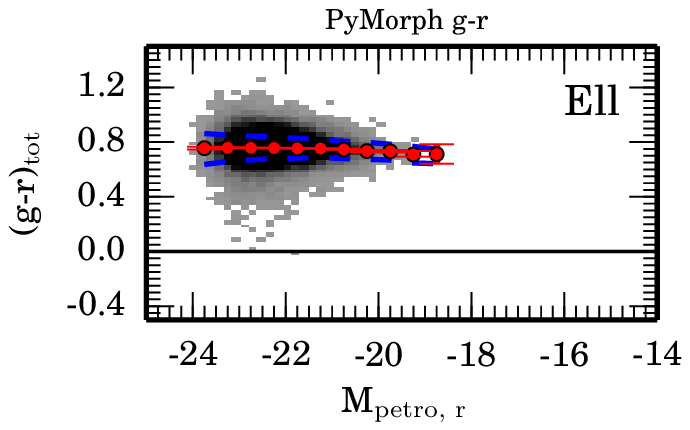}
\includegraphics[width=0.32\linewidth]{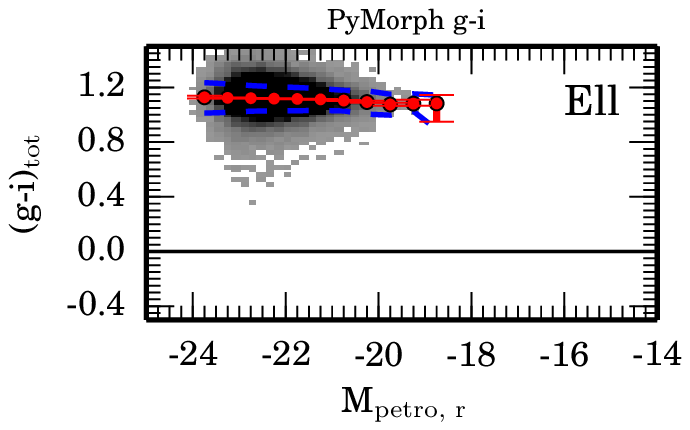}
\includegraphics[width=0.32\linewidth]{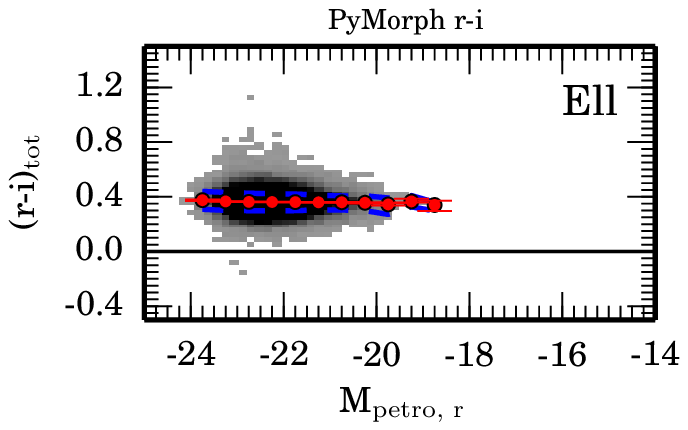}
\includegraphics[width=0.32\linewidth]{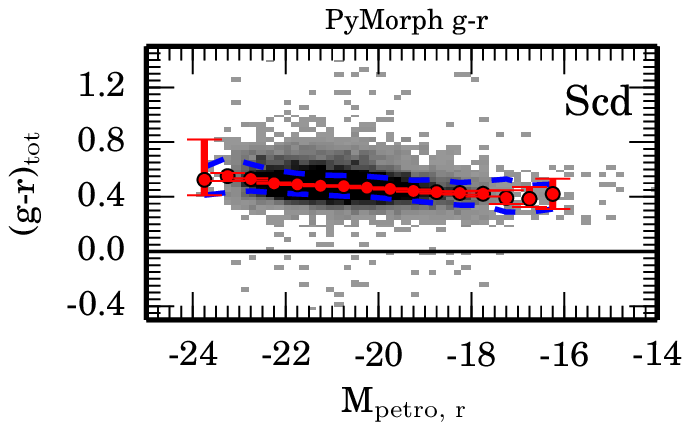}
\includegraphics[width=0.32\linewidth]{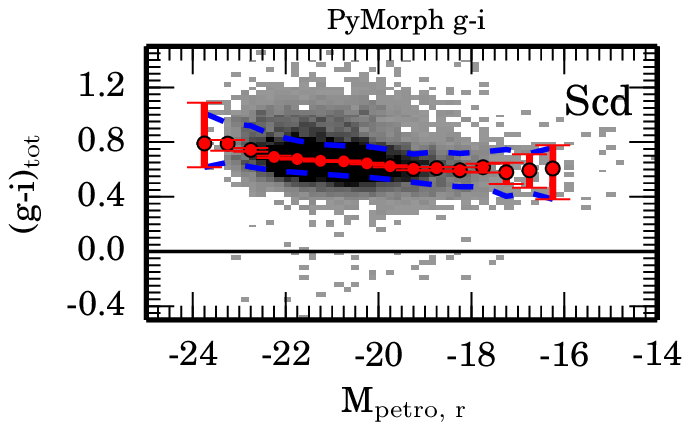}
\includegraphics[width=0.32\linewidth]{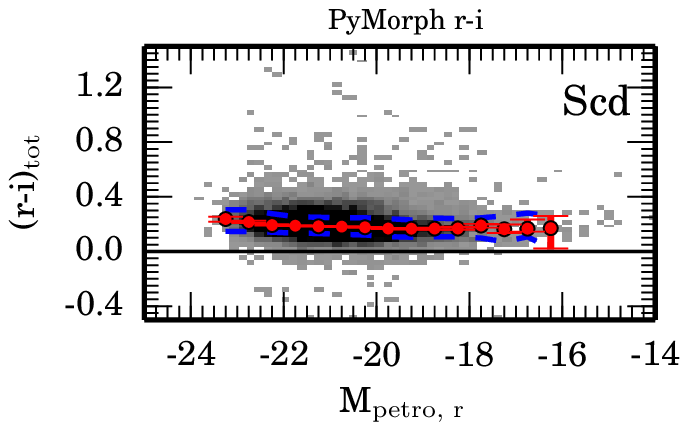}
\caption{\pymorph{} \SerExp{} colours of this work  versus the absolute Petrosian magnitudes of the galaxies. The \pymorph{} aperture colours derived from the \SerExp{} fits of this work are shown for the classes, Elliptical (Ell) and Scd identified in \meertcat{} (rows 1 and 2, respectively). The three combinations of bands {\em g}-{\em r}, {\em g}-{\em i}, and {\em r}-{\em i} in the left-hand, centre, and right-hand columns. 
Median values are plotted in as red symbols with 95 per cent CI as error bars. The 68 per cent contours are shown as blue dashed lines. }\label{fig:color_mag_by_type}	
\end{figure*}

Figure~\ref{fig:color_mag_by_type} shows colours calculated from the light profiles of this work versus the absolute Petrosian magnitudes of the galaxies. The \pymorph{} aperture colours derived from the \SerExp{} fits of this work are shown for the classes Elliptical (Ell) and Scd identified in \meertcat{} (rows 1 and 2, respectively).

There appears to be little change in the total colours across the types. In addition, the scatter is broad enough to make conclusions using the colours difficult.
The typical colour-magnitude relation is visible for Elliptical galaxies (red sequence) and Scd galaxies (blue cloud).
Including S0 and Sab galaxies increase the scatter in these relations as the dominant contribution to the light transitions from the redder Ellipticals to the bluer Scd galaxies.

While we discuss the total colours here, bulge and disc colours can be calculated from our data. Component colours contain even more uncertainty than what is presented here, and the
reader is cautioned to check that the colours are not biased in respect to the particular science being conducted.

\section{Discussion and Use of the Catalogue} \label{sec:discussion}

When using the catalogue, we recommend using the same selection criteria described in Section~8 of \meertcat{}. The format of the data is similar to the format used in \meertcat{}. 
We provide our fits and flags for each of the four models (\Dev{}, \Ser{}, \DevExp{}, and \SerExp) and in each of the three bands. The {\em r}-band data is included in this new data release for completeness and has the aperture magnitudes added into the data. In addition, we provide a preferred catalogue  
for public use. This preferred catalogue has \Ser{} fits for galaxies flagged as pure bulge or pure disc. The remaining galaxies have \SerExp{} fits.

When using the catalogue, we recommend removing galaxies flagged as bad (flag 20) as these galaxies have catastrophically bad estimates of total magnitude and radius. Additional galaxies may be removed depending on how conservative the user seeks to be. The problematic two-component fits (flag 14) or the two-component fits with bulge  S\'{e}rsic index n$=$8 may be used for total magnitude and radius measurements, but the sub-components are not reliable. 

The user should also be aware that we have swapped the bulge and disc components of galaxies with bit 6, 7, or 13 set (which were flagged as inverted profiles in the \SerExp{} fit). These galaxies have B/T inverted and the components reversed relative to the \SerExp{} fit. Therefore, no additional alterations must be made to 
account for the inverted nature of the profile. However, using the ``raw'' fit produced prior to flagging requires swapping bulge and disc parameters and inverting the B/T. This alteration has been done in all catalogues.

We suggest one of these composite samples drawn from the preferred fit catalogue described in the previous paragraph:
\begin{description}
 \item[{\bf The conservative catalogue}] Select all galaxies with final flag bits 11, 12, or 13 set and bulge  S\'{e}rsic index  $<$8. In addition, the user should select galaxies with \SerExp{} final flag bits 1 or 4 set. These galaxies will have B/T of 1 (for bulges; final flag bit 1 set) or a B/T of 0 (for discs; final flag bit 4 set) and the relevant \Ser{} parameters are reported in the catalogue.
 \item[{\bf The intermediate catalogue}] Use the catalogue above plus all galaxies with final flag bit 10 set and bulge  S\'{e}rsic index $=$8.
 \item[{\bf The full catalogue}] Use the catalogue above plus all galaxies with final flag bit 14 set. This is the least restrictive 
 version of the catalogue but may include galaxies with strange, difficult-to-interpret fit parameters.
\end{description}
When using the multi-band data, the selection criteria should be applied across all 3 bands.

While the \Dev{} and \DevExp{} models are not a focus of this work, we recognize that there may be cases in which these 
models are preferred for analysis, especially when comparing to previous work. The breakdown of flags for these models is shown in 
the online Appendix~\ref{app:autoflag}. If the \DevExp{} model is desired for two-component galaxies rather than the \SerExp{} used 
here, the same criteria used to draw the composite samples described in this section may be applied to the \DevExp{} fits. This may
be done at the user's discretion. We point out here that all flagging is carried out in the same manner for the \SerExp{} and 
\DevExp{} models, except where the fits do not allow such treatment (\eg there can be no cases of bulges appearing disc-like 
in the \DevExp{} sample since the S\'{e}rsic index is fixed at 4 in all \DevExp{} fits). The single-component \Dev{} and \Ser{} fits also 
receive the same treatment during flagging.

The data files for this catalogue are available online at \url{http://www.physics.upenn.edu/~ameert/SDSS_PhotDec/}. 

Tables~\ref{tab:modtab1} and \ref{tab:modtab2} describe the format of the data tables released as part of this work. We distribute
the data as a binary table using the \fits{} standard. The first binary extension contains the model independent measurements
for each galaxy (\eg \sextractor{} measurements, the number of fitted neighbours, \etc{}). The following extensions contain the
``best model'' (the combination of \Ser{} and \SerExp{} fits), the \Dev{} model, the \Ser{} model, the \DevExp{} model, and the \SerExp{} model in that order.
These extensions include the fitted values for magnitude, radius, and other parameters as well as the flags described in \meertcat{} (labelled as \texttt{finalflag}, in column 34). 
A separate table containing ra/dec/z information and other identifying information is also available to allow matching between this catalogue and external work.
There are separate files for each of the three bands and the bands are denoted in the name of the file as either ``gband,'' ``rband,'' or ``iband.''

\begin{table*}
\centering
 \begin{tabular}{c c p{9 cm} c }
 \textbf{Column Number} & \textbf{Column Name} & \textbf{Explanation} & \textbf{Data Type} \\ \hline
 0  & SExMag    & the  \sextractor{} magnitude (mag)& float \\
 1  & SExMagErr & the  \sextractor{} magnitude error (mag)& float \\
 2  & SExHrad & the  \sextractor{} half-light radii (arcsec)& float \\
 3  & SExSky    & the  \sextractor{} sky brightness (mag/arcsec$^2$)& float\\
 4  & num\_targets & the number of targets & int\\ 
 5  & num\_neighborfit & the number neighbour sources fitted with \Ser{} profiles& int\\
 6  & AperRad & the  aperture radius (arcsec) used for colour measurements& float \\
 7  &C & the  Concentration& float \\
 8  &C\_err & the  Concentration error& float \\
 9  &A & the  Asymmetry & float \\
 10 &A\_err & the  Asymmetry error& float \\
 11 &S & the  Smoothness & float \\
 12 &S\_err & the  Smoothness error& float \\
 13 &G & the  Gini coefficient & float \\
 14 &M20 & the  M$_{20}$ value & float \\
 15 & extinction & The SDSS-provided galactic extinction (magnitude) & float \\
 16 & dismod   & the calculated distance modulus & float \\
 17 & kpc\_per\_arcsec & the angular scale (kpc/arcsec) & float \\
 18 & Vmax & the Volume used for Vmax corrections (Mpc$^{3}$) & float \\
 19 & SN & the average S/N per pixel inside the half-light radius & float\\ 
 20 & kcorr & k-correction calculated using the SDSS model magnitudes & float 
 \end{tabular}
\caption{Description of columns in the electronic table UPenn\_PhotDec\_nonParam\_g(r)(i)band for the {\em g}, {\em r}, and {\em i}-band data, respectively. 
The data are model-independent measurements fitted by \pymorph{}. 
Problematic data or parameters are replaced with -999.}
\label{tab:modtab1}
\end{table*}
 
\begin{table*}
\centering
 \begin{tabular}{c c p{9 cm} c }
 \textbf{Column Number} & \textbf{Column Name} & \textbf{Explanation} & \textbf{Data Type} \\ \hline
0& m\_tot & total fitted apparent magnitude & float \\
1&m\_aper & the aperture magnitude used measured within the AperRad & float \\
2&BT & the B/T (bulge-to-total light ratio) of the fit & float \\
3&r\_tot & the half-light radius (arcsec) of the total fit & float\\
4&ba\_tot & the axis ratio (semi-minor/semi-major of the total fit & float \\
5&BT\_aper & the B/T (bulge-to-total light ratio) of the fit inside the fixed aperture & float \\
6&xctr\_bulge & the bulge x centre (pixels) & float \\
7&xctr\_bulge\_err & the bulge x centre error (pixels) & float \\
8& yctr\_bulge & the  bulge y centre (pixels) & float \\
9& yctr\_bulge\_err & the  bulge y centre error (pixels) & float \\
10&m\_bulge & the  bulge magnitude & float \\
11& m\_bulge\_err & the  bulge magnitude error& float \\
12& r\_bulge & the  bulge halflight radius (arcsec) & float \\
13& r\_bulge\_err & the  bulge radius error (arcsec)& float \\
14& n\_bulge & the  bulge S\'{e}rsic index & float \\
15& n\_bulge\_err & the  bulge S\'{e}rsic index error & float \\
16& ba\_bulge & the  bulge b/a & float \\
17& ba\_bulge\_err & the  bulge b/a error & float \\
18& pa\_bulge & the  bulge position angle (degrees) & float \\
19& pa\_bulge\_err & the  bulge position angle error (degrees) & float\\
20& xctr\_disc & the  disc x centre (pixels) & float \\
21& xctr\_disc\_err & the  disc x centre error (pixels) & float \\
22& yctr\_disc & the  disc y centre (pixels) & float \\
23& yctr\_disc\_err & the  disc y centre error (pixels) & float \\
24& m\_disc & the  disc magnitude & float \\
25& m\_disc\_err & the  disc magnitude error& float \\
26& r\_disc & the  disc  halflight radius (arcsec) & float \\
27& r\_disc\_err & the  disc radius error (arcsec)& float \\
28& n\_disc & the  disc S\'{e}rsic index & float \\
29& n\_disc\_err & the  disc S\'{e}rsic index error & float \\
30& ba\_disc & the  disc b/a & float \\
31& ba\_disc\_err & the  disc b/a error & float \\
32& pa\_disc & the  disc position angle (degrees) & float \\
33& pa\_disc\_err & the  disc position angle error (degrees) & float \\
34&GalSky    & the  \pymorph{} sky brightness (mag/arcsec$^2$)& float \\
35&GalSky\_err    & the  \pymorph{} sky brightness (mag/arcsec$^2$)& float \\
36&chi2nu & the  $\chi^2/$DOF & float \\
37&finalflag & the primary quality flag described in this work & float \\
38&autoflag & the intermediate, visually calibrated, automated flag described in this work & float \\
39&pyflag & the \pymorph{} run flag  & float \\
40&pyfitflag & the \pymorph{} fit flag & float 
\end{tabular}
\caption{Description of columns in the electronic table 
UPenn\_PhotDec\_Models\_g(r)(i)band for the {\em g}, {\em r}, and {\em i}-band data, respectively. 
The data are the ``best model'', \Dev{}, \Ser{},
\DevExp{}, and \SerExp{} model fit parameters fitted by \pymorph{}. 
Unfit parameters or missing data are replaced with values of -999.}
\label{tab:modtab2}
\end{table*}

\section{Conclusions} \label{sec:conclusion}

The {\em r}-band catalogue of \Dev{}, \Ser{}, \DevExp{}, and \SerExp{} galaxies presented in \meertcat{} constructed for the SDSS DR7 Spectroscopic sample was extended to the {\em g}- and {\em i}-bands. We used 
the \pymorph{} pipeline, including the \sextractor{} and \galfit{} programs, to perform 2D decompositions (see Sections~\ref{sec:data}~and~\ref{sec:fitting}). 
We ised the original flagging system of \meertcat{} where appropriate and applied  
the flagging system to the \SerExp{} fits. 

Depending on the band chosen, we identified 92 to 94 percent of our fitted sample as having reliable total magnitude and half-light measurements. About 33-39 percent of the sample
are two-component galaxies with well-behaved components. An additional $\approx11\%$ may be two-component fits, but with 
difficult-to-interpret components. The remaining 44\% are pure bulge and disk galaxies. We compared the fits across the bands and found fitting to be stable and to exhibit trends with increasing disk contributions toward the {\em g}-band and increasing bulge contributions toward the {\em i}-band. While adjustments for differences in the S/N were considered, simulations 
at S/N values approximating the {\em g}-band fits show that bulge radii have a larger scatter compared to the {\em i}- or {\em r}-bands. Therefore, caution is advised 
when evaluating bulges of the {\em g}-band fits. 

Examination of the \SerExp{} fits and the flagged categories show that the morphological classes assigned in this catalogue using the flags (\ie bulge, disk, or two component) also correlate well across the bands and with absolute magnitude. Aperture colours calculated from these fits show a larger scatter compared to SDSS and have a blueward bias of approxiamtely 0.05 magnitudes relative to SDSS, which appears to be correlated with {\em g}-band measurements. This bias occurs more strongly at the bright
end of the later types (probably due to a larger uncertainty in the g-band since
brighter galaxies are bulge dominated and therefore redder).
The fits also show the usual color magnitude relations when divided into different morphological types
(red sequence and blue cloud).

\section*{Acknowledgements}
We would like to thank our referee for many helpful comments and suggestions. These have greatly improved the paper.

We would also like to thank Mike Jarvis and Joseph Clampitt
for many helpful discussions.

This work was supported in part by NASA grant ADP/NNX09AD02G
and NSF/0908242.

Funding for the SDSS and SDSS-II has been provided by the Alfred P. Sloan
Foundation, the Participating Institutions, the National Science Foundation, the U.S.
Department of Energy, the National Aeronautics and Space Administration, the Japanese
Monbukagakusho, the Max Planck Society, and the Higher Education Funding Council for
England. The SDSS Web Site is http://www.sdss.org/.

The SDSS is managed by the Astrophysical Research Consortium for the
Participating Institutions. The Participating Institutions are the American Museum of
Natural History, Astrophysical Institute Potsdam, University of Basel, University of
Cambridge, Case Western Reserve University, University of Chicago, Drexel University,
Fermilab, the Institute for Advanced Study, the Japan Participation Group, Johns
Hopkins University, the Joint Institute for Nuclear Astrophysics, the Kavli Institute
for Particle Astrophysics and Cosmology, the Korean Scientist Group, the Chinese
Academy of Sciences (LAMOST), Los Alamos National Laboratory, the
Max-Planck-Institute for Astronomy (MPIA), the Max-Planck-Institute for Astrophysics
(MPA), New Mexico State University, Ohio State University, University of Pittsburgh,
University of Portsmouth, Princeton University, the United States Naval Observatory,
and the University of Washington.

\bibliography{bibliography}

\onecolumn

\appendix
\appendixpage

\section{Flag tables for \Dev{}, \Ser{}, and \DevExp{} models}\label{app:autoflag}

\begin{table}
\centering
\begin{tabular}{l l l  p{2cm}  p{2cm}  p{2cm} }
\multicolumn{3}{l}{\textbf{Descriptive Category}} &  \textbf{\% dev,\emph{g}} & \textbf{\% dev,\emph{r}} & \textbf{\% dev,\emph{i}} \\ \hline \hline
\multicolumn{3}{l}{\textbf{Trust Total and Component Magnitudes and Sizes}} & -- & -- & --\\ \hline
& \multicolumn{2}{l}{\textbf{Two-Component Galaxies}} & -- & -- & -- \\
 & & No Flags & -- & -- & -- \\
 & & Good \Ser, Good \Exp\ (Some Flags) & -- & -- & -- \\
 & &Flip Components, n$_{\Ser}<$2 & -- & -- & -- \\ \hline
\multicolumn{3}{l}{\textbf{Trust Total Magnitudes and Sizes Only}} & 96.623 & 97.653 & 96.958\\ \hline
& \multicolumn{2}{l}{\textbf{Bulge Galaxies}} &  96.623 & 97.653 & 96.958\\
& &No \Exp\ Component, n$_{\Ser}>$2&  96.623 & 97.653 & 96.958 \\
& &\Ser\ Dominates Always &  -- & -- & -- \\
& \multicolumn{2}{l}{\textbf{Disk Galaxies}} &  -- & -- & --\\
& & No \Ser\ Component &   -- & -- & --\\
& & No \Exp, n$_{Ser}<$2, Flip Components &   -- & -- & --\\
& & \Ser\ Dominates Always, n$_{\Ser}<$2 &  -- & -- & --\\
& & \Exp\ Dominates Always &   -- & -- & --\\
& & Parallel Components &   -- & -- & --\\
& \multicolumn{2}{l}{\textbf{Problematic Two-Component Galaxies}} &  -- & -- & --\\
& & \Ser\ Outer Only &   -- & -- & --\\
& & \Exp\ Inner Only &   -- & -- & --\\
& & Good \Ser, Bad \Exp, B/T$>=$0.5 &   -- & -- & -- \\
& & Bad \Ser, Good \Exp, B/T$<$0.5 &   -- & -- & -- \\
& & Tiny Bulge, otherwise good &   -- & -- & -- \\ \hline \hline
\multicolumn{3}{l}{\textbf{Bad Total Magnitudes and Sizes}} &  3.377 & 2.347 & 3.042\\ \hline
& \multicolumn{2}{l}{Centering Problems} &  0.292 & 0.338 & 0.311 \\
& \multicolumn{2}{l}{\Ser\ Component Contamination by Neighbors or Sky} &  1.865 & 1.302 & 1.651 \\
& \multicolumn{2}{l}{\Exp\ Component Contamination by Neighbors or Sky} &  -- & -- & -- \\
& \multicolumn{2}{l}{Bad \Ser\ and Bad \Exp\ Components} &  -- & -- & -- \\
& \multicolumn{2}{l}{Galfit Failure} & 0.386 & 0.187 & 0.286 \\
& \multicolumn{2}{l}{Polluted or Fractured} & 1.041 & 0.679 & 0.947 \\
\end{tabular}
 \caption{\Dev{} model table showing flagging breakdown for the g, r, and i bands.}
 \label{tab:dev_flags}
\end{table}

\begin{table}
\centering
\begin{tabular}{l l l  p{2cm}  p{2cm}  p{2cm} }
\multicolumn{3}{l}{\textbf{Descriptive Category}} &  \textbf{\% ser,\emph{g}} & \textbf{\% ser,\emph{r}} & \textbf{\% ser,\emph{i}} \\ \hline \hline
\multicolumn{3}{l}{\textbf{Trust Total and Component Magnitudes and Sizes}} & -- & -- & --\\ \hline
& \multicolumn{2}{l}{\textbf{Two-Component Galaxies}} & -- & -- & -- \\
 & & No Flags & -- & -- & -- \\
 & & Good \Ser, Good \Exp\ (Some Flags) & -- & -- & -- \\
 & &Flip Components, n$_{\Ser}<$2 & -- & -- & -- \\ \hline
\multicolumn{3}{l}{\textbf{Trust Total Magnitudes and Sizes Only}} & 95.502 & 97.380 & 96.612\\ \hline
& \multicolumn{2}{l}{\textbf{Bulge Galaxies}} &  52.177 & 58.378 & 60.972\\
& &No \Exp\ Component, n$_{\Ser}>$2&  52.177 & 58.378 & 60.972 \\
& &\Ser\ Dominates Always &  -- & -- & -- \\
& \multicolumn{2}{l}{\textbf{Disk Galaxies}} &  43.325 & 39.001 & 35.640\\
& & No \Ser\ Component &   -- & -- & --\\
& & No \Exp, n$_{Ser}<$2, Flip Components &   43.325 & 39.001 & 35.640\\
& & \Ser\ Dominates Always, n$_{\Ser}<$2 &  -- & -- & --\\
& & \Exp\ Dominates Always &   -- & -- & --\\
& & Parallel Components &   -- & -- & --\\
& \multicolumn{2}{l}{\textbf{Problematic Two-Component Galaxies}} &  -- & -- & --\\
& & \Ser\ Outer Only &   -- & -- & --\\
& & \Exp\ Inner Only &   -- & -- & --\\
& & Good \Ser, Bad \Exp, B/T$>=$0.5 &   -- & -- & -- \\
& & Bad \Ser, Good \Exp, B/T$<$0.5 &   -- & -- & -- \\
& & Tiny Bulge, otherwise good &   -- & -- & -- \\ \hline \hline
\multicolumn{3}{l}{\textbf{Bad Total Magnitudes and Sizes}} &  4.498 & 2.620 & 3.388\\ \hline
& \multicolumn{2}{l}{Centering Problems} &  0.320 & 0.357 & 0.322 \\
& \multicolumn{2}{l}{\Ser\ Component Contamination by Neighbors or Sky} &  3.012 & 1.618 & 2.036 \\
& \multicolumn{2}{l}{\Exp\ Component Contamination by Neighbors or Sky} &  -- & -- & -- \\
& \multicolumn{2}{l}{Bad \Ser\ and Bad \Exp\ Components} &  -- & -- & -- \\
& \multicolumn{2}{l}{Galfit Failure} & 0.327 & 0.124 & 0.239 \\
& \multicolumn{2}{l}{Polluted or Fractured} & 1.046 & 0.681 & 0.948 \\
\end{tabular}
 \caption{\Ser{} model table showing flagging breakdown for the g, r, and i bands.}
 \label{tab:ser_flags}
\end{table}

\begin{table}
  \centering
 \begin{tabular}{l l l  p{2cm}  p{2cm}  p{2cm} }
\multicolumn{3}{l}{\textbf{Descriptive Category}} &  \textbf{\% devexp,\emph{g}} & \textbf{\% devexp,\emph{r}} & \textbf{\% devexp,\emph{i}} \\ \hline \hline
\multicolumn{3}{l}{\textbf{Trust Total and Component Magnitudes and Sizes}} & 33.077 & 42.223 & 44.524\\ \hline
& \multicolumn{2}{l}{\textbf{Two-Component Galaxies}} & 33.077 & 42.223 & 44.524 \\
 & & No Flags & 22.918 & 30.152 & 31.952 \\
 & & Good \Ser, Good \Exp\ (Some Flags) & 8.526 & 10.701 & 11.180 \\
 & &Flip Components, n$_{\Ser}<$2 & 1.633 & 1.369 & 1.391 \\ \hline
\multicolumn{3}{l}{\textbf{Trust Total Magnitudes and Sizes Only}} & 59.362 & 52.444 & 48.390\\ \hline
& \multicolumn{2}{l}{\textbf{Bulge Galaxies}} &  15.043 & 14.636 & 14.211\\
& &No \Exp\ Component, n$_{\Ser}>$2&  9.610 & 4.646 & 3.398 \\
& &\Ser\ Dominates Always &  5.433 & 9.989 & 10.813 \\
& \multicolumn{2}{l}{\textbf{Disk Galaxies}} &  36.291 & 28.462 & 24.225\\
& & No \Ser\ Component &   34.502 & 25.041 & 20.070\\
& & No \Exp, n$_{Ser}<$2, Flip Components &   -- & -- & --\\
& & \Ser\ Dominates Always, n$_{\Ser}<$2 &  -- & -- & --\\
& & \Exp\ Dominates Always &   1.789 & 3.421 & 4.156\\
& & Parallel Components &   -- & -- & --\\
& \multicolumn{2}{l}{\textbf{Problematic Two-Component Galaxies}} &  8.029 & 9.346 & 9.954\\
& & \Ser\ Outer Only &   4.906 & 5.289 & 5.494\\
& & \Exp\ Inner Only &   0.632 & 0.514 & 0.534\\
& & Good \Ser, Bad \Exp, B/T$>=$0.5 &   0.003 & 0.011 & 0.010 \\
& & Bad \Ser, Good \Exp, B/T$<$0.5 &   0.620 & 0.884 & 1.042 \\
& & Tiny Bulge, otherwise good &   1.868 & 2.648 & 2.874 \\ \hline \hline
\multicolumn{3}{l}{\textbf{Bad Total Magnitudes and Sizes}} &  7.560 & 5.334 & 7.086\\ \hline
& \multicolumn{2}{l}{Centering Problems} &  0.762 & 0.599 & 0.769 \\
& \multicolumn{2}{l}{\Ser\ Component Contamination by Neighbors or Sky} &  1.816 & 1.251 & 1.608 \\
& \multicolumn{2}{l}{\Exp\ Component Contamination by Neighbors or Sky} &  3.887 & 2.788 & 3.868 \\
& \multicolumn{2}{l}{Bad \Ser\ and Bad \Exp\ Components} &  0.634 & 0.177 & 0.148 \\
& \multicolumn{2}{l}{Galfit Failure} & 0.389 & 0.249 & 0.290 \\
& \multicolumn{2}{l}{Polluted or Fractured} & 1.038 & 0.677 & 0.945 \\
\end{tabular}
 \caption{\DevExp{} model table showing flagging breakdown for the g, r, and i bands.}
 \label{tab:devexp_flags}
\end{table}

In this appendix, we provide the breakdown of the flags as a percent of the total sample for the \Dev{}, \Ser{}, and \DevExp{} galaxy fits in each of the three bands. 
These tables are similar in format to Table~\ref{tab:serexp_flags}.
The appendix includes 3 tables

\clearpage
\section{Example Catalogue Data Table}
This appendix includes sample tables of the data released as part of this work. The meaning of the table columns are described in Table~\ref{tab:modtab1} and Table~\ref{tab:modtab2}. The full tables are available on-line in machine-readable format.

\begin{table}
\centering
\begin{tabular}{c c c c c c c c c}
\textbf{SExMag}& \textbf{SExMagErr}& \textbf{SExHrad}& \textbf{SExSky}& 
\ldots& 
\textbf{kpc\_per\_arcsec}& \textbf{Vmax}& \textbf{SN}& \textbf{kcorr}\\ \hline  
17.67 &     0.03 & 2.15 & 20.79 & \ldots &
0.61& 2800960 & 120.20 & 0.05 \\
17.43 &      0.03 & 2.58 & 20.79 &\ldots&
1.47& 58645700& 128.13 & 0.08 \\
17.11 &     0.02 & 3.96 & 20.79 & \ldots&
2.76& 699989000& 123.77 & 0.19 \\
16.93 &     0.02 & 2.75 & 20.79 & \ldots&
1.36& 94692800&181.81 & 0.09 \\
16.70 &      0.02 & 3.29 & 20.79 &\ldots&
1.37& 123920000 &193.85 & 0.13 \\
\ldots &\ldots &\ldots &\ldots &\ldots &\ldots &\ldots 
& \ldots & \ldots \\
\end{tabular}
\caption{Sample of the model-independent measurements described in Table~\ref{tab:modtab1}. The full table has 21 columns and is available in machine-readable format on-line.}
\end{table}

\begin{table}
\centering
\begin{tabular}{c c c c c c c c c}
\textbf{m\_tot}& \textbf{m\_aper}&\textbf{BT}& textbf{r\_tot}&  
\ldots & \textbf{finalflag}& \textbf{autoflag}&  \textbf{pyflag}&
\textbf{pyfitflag}\\ \hline  
17.65 &  17.65&0.21 &   1.76 &   \ldots &  5121 & 2048&62 & 0 \\
17.37 & 17.37&0.05 &   2.28 &    \ldots & 49 &4&62 &     128 \\
17.01 &  17.04&0.14 &   4.11 &  \ldots &  273&131072& 62 &     128 \\
16.81 & 16.86&0.09 &   2.79 &    \ldots & 49&67108868&62 &     384 \\
16.66 & 16.66&0.02 &   3.02 &   \ldots &  49&67108868& 62 &     384 \\ 
\ldots &\ldots &\ldots &\ldots &\ldots &\ldots &\ldots &\ldots &\ldots \\
\end{tabular}
\caption{Sample of the model-dependent measurements for the \SerExp{} fit described in Table~\ref{tab:modtab2}. The full table has 41 columns and is available in machine-readable format on-line. Similar tables are also available for the \Dev{}, \Ser{}, and \DevExp{} fits.}
\end{table}

\end{document}